\documentclass[structabstract]{aa}
\pdfoutput=1
\usepackage{graphicx}
\usepackage{natbib}
\bibpunct{(}{)}{;}{a}{}{,}
\usepackage{txfonts}
\usepackage{multirow}
\usepackage{array}

\begin{document}

\title{Line-of-sight geometrical and instrumental resolution effects on intensity perturbations by sausage modes}

\author{Patrick Antolin \and Tom Van Doorsselaere}
\institute{Centre for mathematical Plasma Astrophysics, Department of Mathematics, KU Leuven\\ Celestijnenlaan 200B, bus 2400, B-3001 Leuven, Belgium\\ \email{[patrick.antolin,tom.vandoorsselaere]@wis.kuleuven.be}}

\titlerunning{Intensity perturbations by sausage modes}
\authorrunning{Antolin \& Van Doorsselaere}

\date{}

\abstract
{Diagnostics of MHD waves in the solar atmosphere is a topic which often encounters problems of interpretation, due partly to the high complexity of the solar atmospheric medium. Forward modeling can significantly guide interpretation, bridging the gap between numerical simulations and observations, and increasing the reliability of mode identification for application of MHD seismology.}
{Determine the characteristics of the fast MHD sausage mode in the corona on the modulation of observable quantities such as line intensity and spectral line broadening.  Effects of line-of-sight angle, and spatial, temporal and spectral resolutions are considered.}
{We take a cylindrical tube simulating a loop in a low-$\beta$ coronal environment with an optically thin background, and let it oscillate with the fast sausage mode. A parametric study is performed.} 
{Longitudinal structuring of the intensity modulation is obtained, set by the nodal structure of the radial velocity. The modulation is strongly dependent on the contribution function of the spectral line. Under the assumption of equilibrium ionisation, the intensity variation can be very low ($\lesssim4~\%$ for \ion{Fe}{ix}~171) or significant ($35~\%$ for \ion{Fe}{xii}~193). Most of this variation disappears when considering the radiative relaxation times of the ions, due to the fast timescales of the sausage mode in the corona. Regardless of the ionisation state of the plasma, the variation of spectral line broadening can be significant, even for low intensity modulation. The nature of this broadening is not thermal but is mostly turbulent. This places spectrometers in clear advantage over imaging instruments for the detection of the sausage mode. The modulation of all quantities can considerably decrease with the line-of-sight angle with respect to the perpendicular to the tube axis. The spatial and temporal resolution are the main factors affecting modulation, erasing longitudinal structuring when these are on the order of the mode's wavelength or the mode's period, placing high constraints on instrumentation. Significant variability in all quantities can still be obtained when viewing at an angle of up to $30^{\circ}$, with pixel size resolutions up to 1/3 of the mode's wavelength, or temporal resolution of one fifth of the mode's period. Modulation is only weakly dependent on spectral resolution due to the mode's inherent symmetry.}
{}

\maketitle

\section{Introduction}

During the last decade solar physics has witnessed a boom of observational reports on wave-like behaviour in all layers of the solar atmosphere \citep[see the recent review by][]{DeMoortel_Nakariakov_2012RSPTA.370.3193D}. This is due, on the one hand, to the advances in imaging and spectroscopic instruments, allowing higher spatial, temporal and spectral resolutions, which have allowed the proper visualization of MHD waves with a rather large range of properties in terms of amplitudes, periods and speeds. On the other hand, theoretical progress points to the strong potential of waves for explaining the heating demands of the chromosphere and the corona in the observed spatial and time scales, as well as the important role they can play for the indirect estimation of local physical parameters of the plasma \citep[MHD seismology, ][]{Roberts_1984ApJ...279..857R,Uchida_1970PASJ...22..341U, Aschwanden_2004psci.book.....A, Taroyan_2009SSRv..tmp...24T}. This of course applies to all physical parameters dictating the behaviour of these waves: the magnetic field strength \citep{Nakariakov_1999Sci...285..862N, Pascoe_etal_2007AA...461.1149P}, the magnetic field geometry \citep{Verth_Erdelyi_2008AA...486.1015V, Ruderman_2008ApJ...686..694R, VanDoorsselaere_etal_2009SSRv..149..299V}, the density inhomogeneity \citep{Aschwanden_Nightingale_2003ApJ...598.1375A, Goossens_2002AA...394L..39G, Goossens_2008AA...484..851G, Arregui_2007AA...463..333A}, the temperature structure, the response of the plasma to energy injection \citep[polytropic index, ][]{VanDoorsselaere_etal_2011ApJ...727L..32V}, the plasma $\beta$ parameter \citep{VanDoorsselaere_2011ApJ...740...90V}, and the transport coefficients \citep{Nakariakov_Verwichte_2005LRSP....2....3N}. Therefore, much emphasis has been put on the search for all kinds of MHD waves, each contributing in a more or less specific way to the measurement of a specific quantity. 

 
The fast MHD sausage mode ($m=0$ mode), which we consider in this study, is compressible, characterised by axi-symmetric radial displacements of the plasma, having magnetic pressure and gas pressure acting in phase as restoring forces. Thus, this mode does not perturb the main axis of loops but its cross-section, producing intensity variations and spectral line broadening.

The MHD fast sausage mode has been first described analytically by \citet{Rosenberg_1970AA.....9..159R}, \citet{Zaitsev_Stepanov_1975AA....45..135Z}, \citet{Roberts_1984ApJ...279..857R}, and \citet{Cally_1986SoPh..103..277C}. Most observations of oscillatory phenomena interpreted as sausage modes have been made in events associated to solar flares, observed with radiometers in the microwave range. These have been most frequently reported using the Nobeyama Radioheliograph \citep[NoRH, ][]{Nakajima_etal_1994IEEEP..82..705N}, with a cadence of 0.1~s and a spatial resolution of $5\arcsec-10\arcsec$ observing in microwave frequencies of 17~GHz and 34~GHz \citep{Nakariakov_2003AA...412L...7N, Melnikov_etal_2005AA...439..727M, Reznikova_2007ARep...51..588R, Inglis_etal_2008AA...487.1147I}. The RHESSI spectrometer \citep{Lin_etal_2002SoPh..210....3L}, with a time resolution of 4~s and a spatial resolution of $2.3\arcsec$ in the $3-100~$keV range, has also been able to observe modulation characteristic of sausage modes in the hard X-rays range \citep{Zimovets_Struminsky_2010SoPh..263..163Z}. In observations with the LYRA radiometer aboard PROBA2, which includes both chromospheric and coronal channels, \citet{VanDoorsselaere_2011ApJ...740...90V} report two modes of oscillation in a single flaring loop, which were interpreted as slow and fast sausage modes. Few reports also exist in EUV lines with the EIS spectrometer of Hinode \citep{Kitagawa_Yokoyama_2010ApJ...721..744K}, with the Atmospheric Imaging Assembly of SDO \citep{Su_2012ApJ...755..113S} and with SECIS \citep{Williams_etal_2002MNRAS.336..747W,Williams_etal_2001MNRAS.326..428W}. Similarly, few reports also exist in cooler (chromospheric and photospheric) lines, with ARIES \citep{Srivastava_2008MNRAS.388.1899S}, with the Solar Optical Telescope of Hinode \citep{Fujimura_Tsuneta_2009ApJ...702.1443F}, and with the ROSA instrument \citep{Morton_2011ApJ...729L..18M, Jess_2012ApJ...744L...5J, Morton_1012NatCommun...3...1315}. 

Most observations of fast sausage modes are linked to Quasi-Periodic Pulsations (QPPs) commonly observed during solar flares in microwave and soft and hard X-ray ranges \citep[see][for reviews on the subject]{Nakariakov_2005LRSP....2....3N, DeMoortel_Nakariakov_2012RSPTA.370.3193D}. As shown by \citet{Mossessian_Fleishman_2012ApJ...748..140M}, not only the sausage mode but also the kink and the torsional Alfv\'{e}n mode can lead to modulation of the source parameters (e.g. energetic electron distribution, magnetic field strength and column thickness along the line-of-sight) resulting in gyrosynchrotron radiation. 

\citet{Nakariakov_2003AA...412L...7N} have shown that the period of the sausage mode in the corona is mostly determined by the length of the hosting structure and the external Alfv\'{e}n speed, thus leading to very short periods. They further show that non-leaky sausage modes occur in sufficiently thick and dense loops, thus greatly constraining their existence in the solar atmosphere \citep{Nakariakov_Hornsey_Melnikov_2012}. It is therefore not surprising that most of the reported sausage modes in the literature come from observations with high cadence instruments, mostly radiometers, and are mostly associated with flaring loops, where densities are high. However, by performing 2D MHD simulations, \citet{Pascoe_etal_2007AA...461.1149P} have investigated the existence of sausage modes in long coronal loops with weak density contrast by extending the wave numbers into the leaky regime. In flaring conditions, they found sausage modes with periods from 5~s to 60~s of detectable quality in loops of length up to 60~Mm, thus supporting the interpretation from observations of QPPs. In this regime the period of the sausage mode is determined by the length of the loop and the external Alfv\'{e}n speed, therefore providing a seismological way to measure the latter and the external magnetic field. Very few reports exist, however, of sausage modes in long coronal (non-flaring) loops, a fact that may be linked either to MHD radiation (leaky modes) leading to a poor quality factor for such modes, or to the lack of adequate temporal and spatial resolutions.


In a 3D MHD model \citet{DeMoortel_Pascoe_2012ApJ...746...31D} analyze the effects of line-of-sight integration on energy budget estimates in multi-stranded coronal loops subject to propagating kink modes. The modes strongly couple to azimuthal Alfv\'{e}n waves, to which all energy is eventually transferred, resulting in kinetic energy estimates from Doppler shifts 1 to 2 orders of magnitude lower than the true wave energy. This suggests that most of the energy could be observed in spectral line broadening.

\citet{Cooper_2003AA...397..765C, Cooper_2003AA...409..325C} investigate analytically with a 2D model the  modulation of intensity from propagating kink and sausage modes based on variation of the angle between the line-of-sight and the emitting structure. They further consider the effect from curvature, and apply the results to explain propagating intensity perturbations along a loop observed with SECIS \citep{Williams_etal_2001MNRAS.326..428W, Williams_etal_2002MNRAS.336..747W}.

\citet{Gruszecki_etal_2012AA...543A..12G} have performed 3D numerical simulations of (trapped and leaky) sausage modes oscillating in a cylinder, and study the resulting modulation of the intensity along a ray perpendicular to the axis of the cylinder. They consider the variation of the pixel size, thus simulating different spatial resolution during observations, and find that intensity variation essentially disappears for resolutions on the order of the mode's wavelength.

These previous works calculate the intensity by integrating the density squared along the line-of-sight, and therefore ignore the radiative response of the plasma and effects from instrumental resolution. Based on numerical simulations, synthetic observations of TRACE channels and Hinode/EIS by \citet{DeMoortel_Bradshaw_2008SoPh..252..101D} illustrate the `ill-posed' intensity inversion problem, which can even alter the period of oscillations. Similarly, synthetic observations of AIA channels of SDO by \citet{Martinez-Sykora_etal_2011ApJ...743...23M} show that significant contribution to the emerging intensity can come from non-dominant ions of a specific channel. These results imply that not only density but also ionisation balance and emissivity need to be taken into account for forward modelling intensities.


The application of MHD seismology strongly relies on the capacity to pinpoint the nature of the wave in observations. However, as analytical, numerical and observational studies show, mode identification is far from a trivial matter. In this work we choose to use a simple approach to characterise the observable properties of the MHD fast sausage mode in the corona. This approach can be used as a platform for constructing more complex models or can help understand more complex numerical results, such as those coming from 3D numerical simulations. We assume that we have a cylindrical structure in a low-$\beta$ coronal environment, which offers significant density contrast with the ambient plasma, and hosts the fast sausage mode. By synthesising the emerging intensity in specific coronal lines using the CHIANTI atomic database \citep{Dere_1997AAS..125..149D,Dere_etal_2009AA...498..915D} we investigate the effects from varying line-of-sight angles, as well as effects from spatial, temporal and wavelength resolutions on the modulation of observable quantities. By doing so we are able to predict the detectability of this mode with current instruments, as well as incoming instruments in the near future.

The paper is structured as follows. In section~\ref{modeldescrip} we describe the different models considered, in section~\ref{results} we present the results from two different point of views: that of imaging instruments, and that of instruments with spectrometric capabilities. In section~\ref{discuss} we discuss the results and conclude in section~\ref{conclus}.

\section{Models}\label{modeldescrip}
\subsection{Variation of thermodynamic quantities}\label{varther}
In the forward modeling approach, the task of identifying MHD modes in the solar corona starts by implementing these modes in realistic numerical simulations. For the latter, it is necessary first to identify the theoretical observational signatures from such modes, free of nonlinear effects. In this work we take such an approach and consider the simplest model of an axisymmetric plasma cylinder of radius $R$ embedded in a low-$\beta$ coronal environment, oscillating with the sausage mode, i.e. the axisymmetric $m=0$ mode described in \citet{Edwin_Roberts_1983SoPh...88..179E}. In this study we ignore the effects of gravity, field-line curvature and twist. The variation of the thermodynamic quantities are found by linearizing the perturbed ideal magnetohydrodynamic (MHD) equations about the magnetostatic equilibrium. In a cylindrical coordinate system $(r,\phi,z)$ we have:
\begin{eqnarray}
\frac{\partial\rho}{\partial t}&=&-\rho_{0}\nabla\cdot \textbf{v} \\
\rho\frac{\partial \textbf{v}}{\partial t} &=& -\nabla p +\frac{1}{\mu}(\nabla\times\textbf{B})\times\textbf{B$_{0}$} \\
\frac{\partial\textbf{B}}{\partial t} &=& \nabla\times(\textbf{v}\times\textbf{B$_{0}$}) \\
\frac{\partial p}{\partial t} &=& \frac{\gamma p_0}{\rho_0}\frac{\partial\rho}{\partial t}
\end{eqnarray}
where $\textbf{v}=(v_{r},v_{\phi},v_{z})$ is the perturbation velocity, $\rho$, $p$, $\textbf{B}$ are the perturbed mass density, gas pressure and magnetic field, and $\rho_0$, $p_0$, $\textbf{B$_0$}=(0,0,B_{z,0})$ correspond to the unperturbed quantities (with $\mu$ the magnetic permeability and $\gamma=5/3$ the ratio of specific heats). In the case of a sinusoidal perturbation leading to a standing mode we have $\nabla\cdot\textbf{v}=-{\cal R}(r)A\cos(\omega t)\sin(kz)$, where $A$ is the total amplitude of the perturbation and $k=\pi N/L_{0}$ is the longitudinal wavenumber (with $N$ the longitudinal mode number and $L_{0}$ the length of the cylinder). In the case of a sausage mode, $\cal{R}$, the radial dependence of the divergence of $\textbf{v}$ must satisfy
\begin{equation}
\frac{\mbox{\rm{d}}^{2}\cal{R}}{\mbox{\rm{d}}r}+\frac{1}{r}\frac{\mbox{\rm{d}}\cal{R}}{\rm{d}r}-\kappa^{2}{\cal R}=0.
\end{equation}
Here, $\kappa^{2}=\frac{(k^{2}c_{s}^{2}-\omega^{2})(k^{2}v_{A}^{2}-\omega^{2})}{(c_{s}^{2}+v_{A}^{2})(k^{2}c_{t}^{2}-\omega^{2})}$, $c_{t}^{2}=c_{s}^{2}v_{A}^{2}/(c_s^{2}+v_{A}^{2})$, where $v_{A}=B_{z,0}/\sqrt{\mu\rho}$ is the Alfv\'{e}n speed and $c_{s}=\sqrt{\gamma p/\rho}$ is the sound speed. $\omega$ satisfies the dispersion relation: 
\begin{equation}
\frac{\kappa_{e}}{\rho_{e}(k^{2}v_{A_{e}}^{2}-\omega^{2})}\frac{K_{0}'(\kappa_{e} R)}{K_{0}(\kappa_{e} R)}=\frac{\kappa_{i}}{\rho_{i}(k^{2}v_{A_{i}}^{2}-\omega^{2})}\frac{J_{0}'(\sqrt{-\kappa_{i}^{2}} R)}{J_{0}(\sqrt{-\kappa_{i}^{2}} R)},
\end{equation}
where $K_{0}$ and $J_{0}$ are the 0-th order Bessel functions. The perturbed variables of interest become:
\begin{eqnarray}\label{standing}
\rho &=& \frac{A\rho_{0} \cal{R}}{\omega}\sin(\omega t)\sin(kz) \\
p &=& \frac{A\gamma p_{0}\cal{R}}{\omega}\sin(\omega t)\sin(kz) \\
T &=& \frac{A(\gamma-1)T_{0}\cal{R}}{\omega}\sin(\omega t)\sin(kz) \\
v_{r} &=& \frac{A(c_{s}^{2}+v_{A}^{2})\left(1-\frac{k^{2}c_{t}^{2}}{\omega^{2}}\right)\frac{\mbox{\rm{d}}\cal{R}}{\mbox{\rm{d}}r}}{\omega^2-k^{2}v_{A}^{2}}\cos(\omega t)\sin(kz) \\
v_{\phi} &=& 0 \\
v_{z} &=& \frac{Akc_{s}^{2}\cal{R}}{\omega^{2}}\cos(\omega t)\cos(kz),
\end{eqnarray}
The unperturbed thermodynamic quantities are linked through the ideal gas law for a fully ionized hydrogen gas $p_{0}=2\frac{k_{B}}{m_{H}}\rho_{0} T_{0}$. 

\begin{figure}
\centering
\includegraphics[width=9cm]{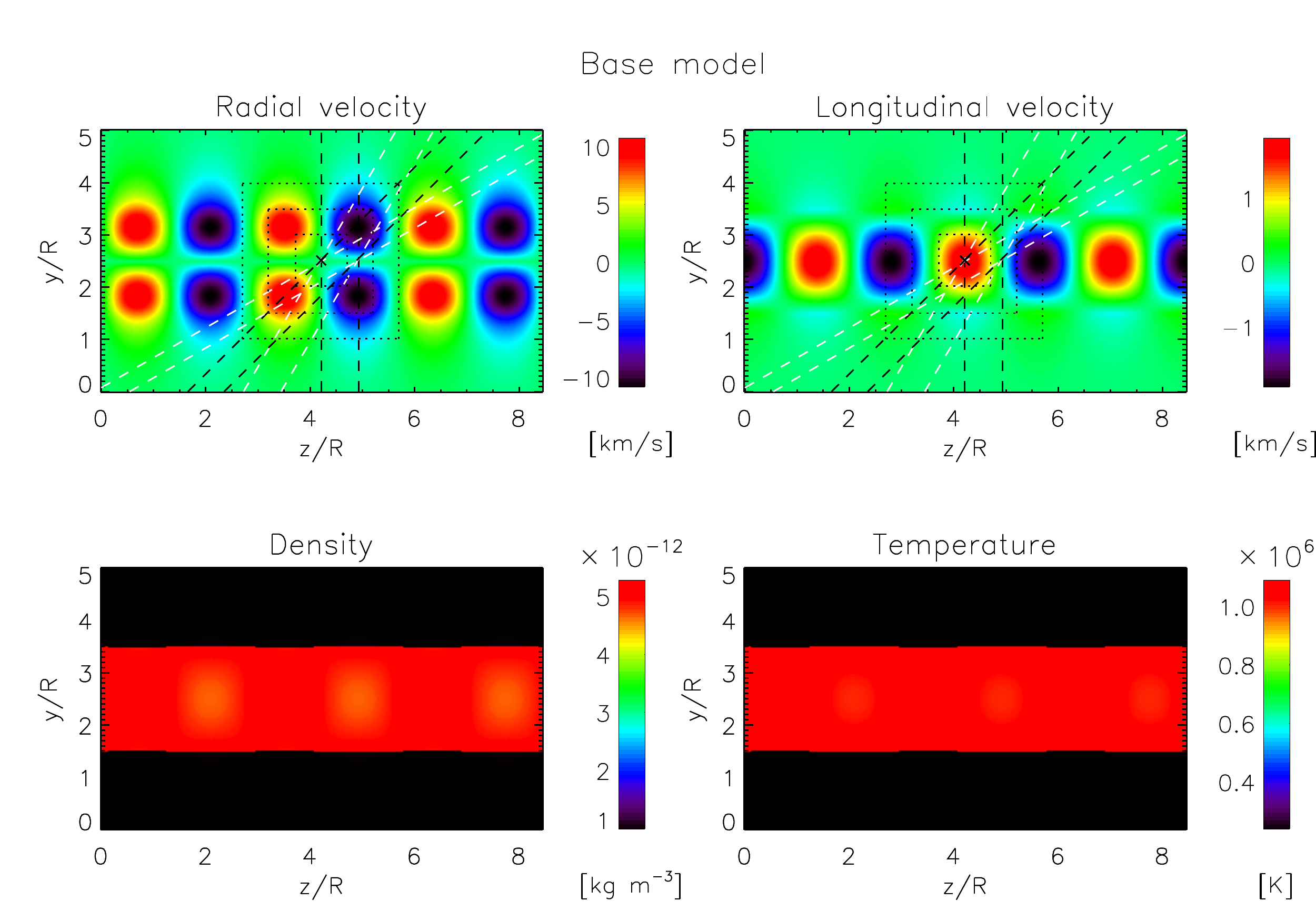}
\caption{Maps of radial velocity (\textit{top-left}), longitudinal velocity (\textit{top-right}), density (\textit{bottom-left}), and temperature (\textit{bottom-right}) along the middle of the cylinder for the \textit{base model - 171} (see Sect.~\ref{models} for more details). The snapshot is taken at 1/3 of the mode period. The black and white dashed lines in the top panels denote the 4 different line-of-sight angles considered: $0^{\circ}, 30^{\circ}, 45^{\circ}$ and $60^{\circ}$. For each angle two rays along the line-of-sight are plotted, one passing through a node of the radial velocity  (left ray of the pair) and one passing through an anti-node of the same quantity (right ray of the pair). The different pixel sizes considered are shown centred at a node: `0R' (=25~km), `1R' = 1~Mm, `2R' = 2~Mm, `3R' = 3~Mm. In the base model, 1R, 2R and 3R correspond roughly to $\ell/3,2\ell/3$ and $\ell$, respectively, where $\ell$ denotes the wavelength of the mode.}
\label{fig1}
\end{figure}

\begin{figure}
\centering
\includegraphics[width=9cm]{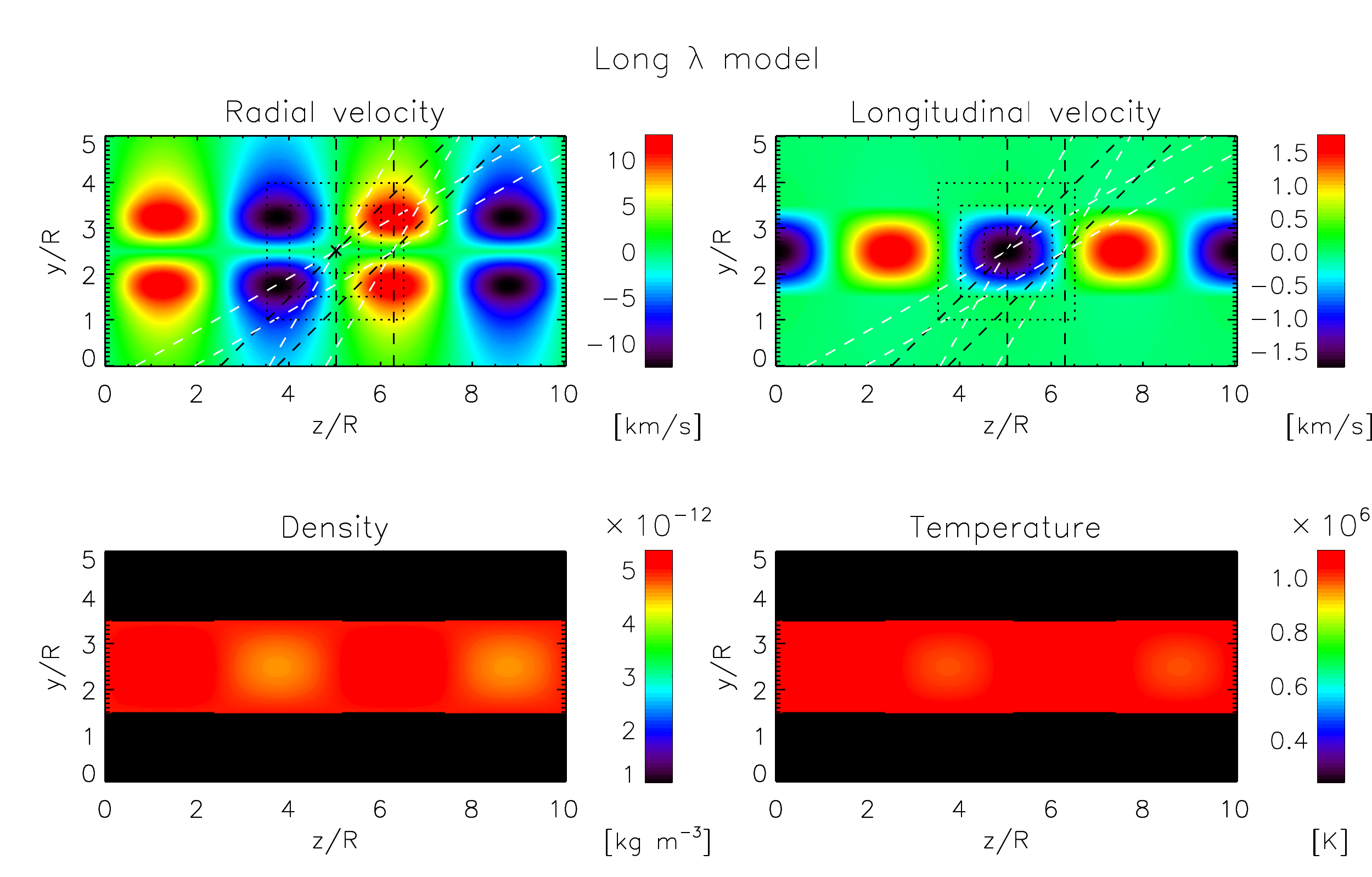}
\caption{Same as Fig.~\ref{fig1} but for the long~$\lambda$ model. See Sect.~\ref{models} for more details.}
\label{fig2}
\end{figure}

\subsection{Base model}\label{base}
Once we know the analytical solution for the thermodynamic quantities we produce a 3D data cube of a cylinder oscillating with the sausage mode. All the models considered in this work are variations of a `base model'. For comparison purposes we choose our base model to be similar to the simulation model by \citet{Gruszecki_etal_2012AA...543A..12G}. Accordingly, we take a discontinuous density profile with a mass density contrast $\rho_{i}/\rho_{e}=5$ and a magnetic field ratio $B_{i}/B_{e}=0.89$ between the interior and the exterior of the unperturbed cylinder. Correspondingly, the Alfv\'{e}n speeds inside and outside the cylinder oscillate around $350~\mathrm{km~s}^{-1}$ and $900~\mathrm{km~s}^{-1}$, respectively. The plasma $\beta$ parameter is at all times lower than 0.3. The internal and external temperature for the unperturbed cylinder are set in this model to $1.05\times10^{6}~\mathrm{K}$ and $2.4\times10^{5}~\mathrm{K}$. For further analogy with \citet{Gruszecki_etal_2012AA...543A..12G}, we choose the wavenumber in the base model as $kR=2.24$, where $R=1~\mathrm{Mm}$ denotes the radius of the cylinder. This results in a wavelength of $\approx2.8~\mathrm{Mm}$. The amplitude $A$ of the standing mode (see Eq.~\ref{standing}) is chosen such as to generate velocity perturbations in agreement with typical values deduced from spectral line broadening in the corona \citep{Chae_etal_1998ApJ...505..957C, Peter_2010AA...521A..51P}. Accordingly, we choose $A=0.011$ \citep[similar to the value used in][]{Gruszecki_etal_2012AA...543A..12G}.

Typical sizes for plasma inhomogeneities in the corona are still not well known. This has lead to several debates concerning the concept of loops as monolithic or rather multi-stranded structures in the corona. Currently, state-of-the-art coronal imaging devices such as SDO/AIA with instrumental point spread functions (PSF) on the order of $1 - 1.2$~Mm are providing increasing observational evidence for a family of structures with sizes not resolved by the telescopes \citep{Brooks_2012ApJ...755L..33B}. Higher resolution imaging can be achieved with chromospheric instruments through tracing of coronal rain in loops, which has provided evidence for strand like structures on the order of a few hundred km \citep{Antolin_Rouppe_2012ApJ...745..152A}. While these works reveal sub-arcsecond density structure providing support for multi-stranded loop models, it is reasonable to think that the collection of strands composing the loop (defining the main waveguide) is actually much larger and behave collectively, agreeing with our choice of 1~Mm for the radius of our loop. Such families of neighbour strands traced by coronal rain indeed have several Mm in diameters, as seen in \citet{Antolin_Verwichte_2011ApJ...736..121A} or\citet{Antolin_Rouppe_2012ApJ...745..152A}. Furthermore, it has been shown that the characteristics of a sausage mode are rather insensitive to the presence of cross-sectional inhomogeneities in loops  \citep{Pascoe_Nakariakov_2007SoPh..246..165P, Pascoe_Nakariakov_2009AA...494.1119P}.

Numerically, our numerical model consists of a $204\times204\times N_{z}$ grid. $N_{z}$ denotes the number of points in the $z$ direction and is chosen so that the different line-of-sight rays for the intensity calculation (explained in Sect.\ref{spectrum}) traverse the same cross-sectional area in the model. In the base model, three wavelengths are enough, leading to $N_{z}=342$ points. The spatial resolution in $x$, $y$ and $z$ axes is the same: about 25~km. The fast sausage mode in the corona produces very low cross-sectional area variation, which for some of the considered models leads to very low intensity variations. To correctly calculate the variations in those models we have also considered a numerical model with 10 times the resolution in each direction. Correspondingly, all results related to intensity modulation are obtained with the high resolution models. All other results do not vary significantly between the low and high spatial resolution models.

Figure~\ref{fig1} shows a snapshot of our base model, representing a cut along the middle axis of the cylinder. The sausage mode creates in-phase perturbations between the density and the magnetic field, and anti-phase perturbations between the density and the radius. Furthermore, the radial and longitudinal velocities are $\pi/4$ out of phase with the density. The snapshot in Fig.~\ref{fig1} is taken at a time of 1/3 of the period $P$, which, in this model, corresponds to $t\approx4.6~\mathrm{s}$. Positive velocity values correspond to outflow towards the axis of the cylinder.  From Fig.~\ref{fig1} it can be noticed that the perturbation of longitudinal velocity is about 1 order of magnitude smaller than the radial perturbation. They have maximum values of about $19~\mathrm{km~s}^{-1}$ and $3.4~\mathrm{km~s}^{-1}$, respectively. At the location of an anti-node of the radial velocity, density oscillates about 7~\% around a mean value of $4.9\times10^{-12}~\mathrm{kg~m}^{-3}$, and temperature oscillates about 5~\% around a mean value of $1.048\times10^{6}~\mathrm{K}$. 

\subsection{Calculation of synthetic spectra}\label{spectrum}

For the calculation of synthetic spectra we choose different viewing angles on the structure, simulating different line-of-sight configurations. Denoting by $\theta$ the angle between the line-of-sight and the perpendicular to the axis of the cylinder, we consider the cases: $\theta=0^{\circ}, 30^{\circ}, 45^{\circ}, 60^{\circ}$, shown in white or black dashed lines in Fig.~\ref{fig1}. For each line-of-sight, we define ray 1 and ray 2 as two particular rays crossing the cylinder at a node and an anti-node of the radial velocity, respectively (unless stated otherwise, the terms `node' and `anti-node' will refer to the radial velocity amplitude throughout the manuscript). For simulating observations it is futher important to consider the spatial, temporal and spectral resolution of the instrument at hand. For each ray we consider 4 different resolution `pixels': `0R' (0 radius, equal to the spatial resolution of our model, i.e. 25 km), `1R' (1 radius~$\approx1/3\ell$, where $\ell$ is the wavelength of the mode in the base model), `2R' (2 radii $ \approx2/3\ell$) and `3R' ($\approx\ell$). These pixel sizes correspond to the dotted squares in Fig.~\ref{fig1}. A pixel of 1~R would correspond roughly to the size of the instrumental PSF for the 171 and 193 channels of the AIA instrument of SDO \citep{Grigris_etal_psf}. Taking into account that the timescale of the sausage mode in a coronal structure such as this one is on the order of seconds (the period $P$ of the mode in the base model is about 5~s), which is very low with respect to the current cadence of available instruments (at most 1~s for imaging instruments, in general 12~s), we consider the following temporal resolutions: `0s' (equal to the temporal resolution in our model, i.e. 0.15~s), `1~s' ($\simeq P/5$) and `2~s' ($\simeq3P/7$). The spectral resolution in our numerical model is set at $1.4~\mathrm{m\AA}$ for the \ion{Fe}{ix}~171 line and $1~\mathrm{m\AA}$ for the \ion{Fe}{xii}~193 line, leading to a spectral sampling of 48 points across 99~\% of the spectrum (details about the selected lines are given below). We also consider coarser spectral resolutions of $\Delta\lambda=4.2$~m$\AA$ and $\Delta\lambda=8.5$~m$\AA$, leading to samplings of 16 and 8 points respectively.

The quantity that we are mostly interested in synthesising is the specific intensity for optically thin (coronal) plasma along a given line-of-sight:
\begin{equation}\label{intens}
I_{\lambda}=A_{\mathrm{b}}g_{\mathrm{eff}}f_{\lambda_{0}}\int G_{\lambda_{0}}(T,n_{e})n_{e}^{2}dl,
\end{equation}
where $A_{\mathrm{b}}$ is the (coronal) abundance of the element (denoted by its wavelength at line centre $\lambda_{0}$) relative to hydrogen, $g_{\mathrm{eff}}$ is the Gaunt factor, $f_{\lambda_{0}}$ is the oscillator strength of the line, $G_{\lambda_{0}}(T,n_{e})$ is a mostly temperature dependent function characterising the response of the plasma for that particular line, $n_{e}$ is the electron number density, and where the integral is evaluated along a ray crossing the emitting plasma ($dl$ is an element of length along the line-of-sight). The $G_{\lambda_{0}}(T,n_{e})$ function depends mainly on the ion population fraction, which can be obtained by solving the time-dependent ion balance equations. While such a treatment would provide a thorough analysis of the radiative response of the plasma, it would take us outside the scope of this paper \citep[such a treatment has been followed by, e.g.][]{Bradshaw_Mason_2003AA...407.1127B}. Here, we will assume that the plasma is at all times in equilibrium ionisation and we will estimate the effects of non-equilibrium on the observable quantities.

The purpose of this work is not to determine exactly the absolute value of line intensities, but rather to analyze the effect on the intensities from temperature, density and especially the geometry (the line-of-sight rays). Hence, we concentrate here on the quantities that compose the integrand in Eq.~\ref{intens}. The calculation of the integral in Eq.~\ref{intens} for all rays crossing the domain is a computationally expensive procedure that can be considerably sped up with the following techniques. For the calculation of $G_{\lambda_{0}}(T,n_{e})$ we produce a look-up table depending on temperature and density using the CHIANTI atomic database \citep{Dere_1997AAS..125..149D, Dere_etal_2009AA...498..915D}. For the latter we choose the chianti collisional ionisation equilibria file (\texttt{chianti.ioneq}) and coronal abundances \citep[\texttt{sun\_coronal.abund} file, from][]{Feldman_1992PhyS...46..202F}. The look-up table is composed by a $2000\times100$ points uniform sampling in the electron number density $n_{e}$ - temperature $T$ domain (respectively), where $\log n_{e}\in[8,10]$ and $\log T\in[4.95,6.5]$, which covers the values obtained in our numerical model. Since the $G$ function is mostly dependent upon temperature the values of the table are only interpolated to the temperature values of each pixel in our model. We produce histograms of the line-of-sight velocity choosing a bin size of 0.5~km~s$^{-1}$ (much smaller than the plasma motions produced by the perturbation) and work with the resulting bins instead of pixel-by-pixel. For each resulting bin we produce a second histogram of the emissivity $G_{\lambda_{0}}(T,n_{e})n_{e}^{2}$ with the data points within the bin. Thanks to the symmetry in our model the bin number necessary for high accuracy for this second histogram is lower than 10000. The spectral line is then calculated for each emissivity bin by assuming a Gaussian distribution for the plasma. The maximum of the Gaussian is given by the value of the emissivity bin. The width of the Gaussian is set by the thermal broadening, given by the logarithmic average temperature of the pixels within the emissivity bin. The displacement of the Gaussian's centre is set by the bin value of the velocity histogram. The last step consists in integrating along the line-of-sight the Doppler shifted emission. This involves an interpolation from the values of the cartesian grid of the Doppler shifted emission to equally spaced points along the line-of-sight rays. With the outlined method the calculation time of the synthetic spectrum for one line for 4 different line-of-sight angles is 8 to 10 times faster than with a normal pixel-by-pixel method. 

\begin{figure}
\centering
\includegraphics[width=9cm]{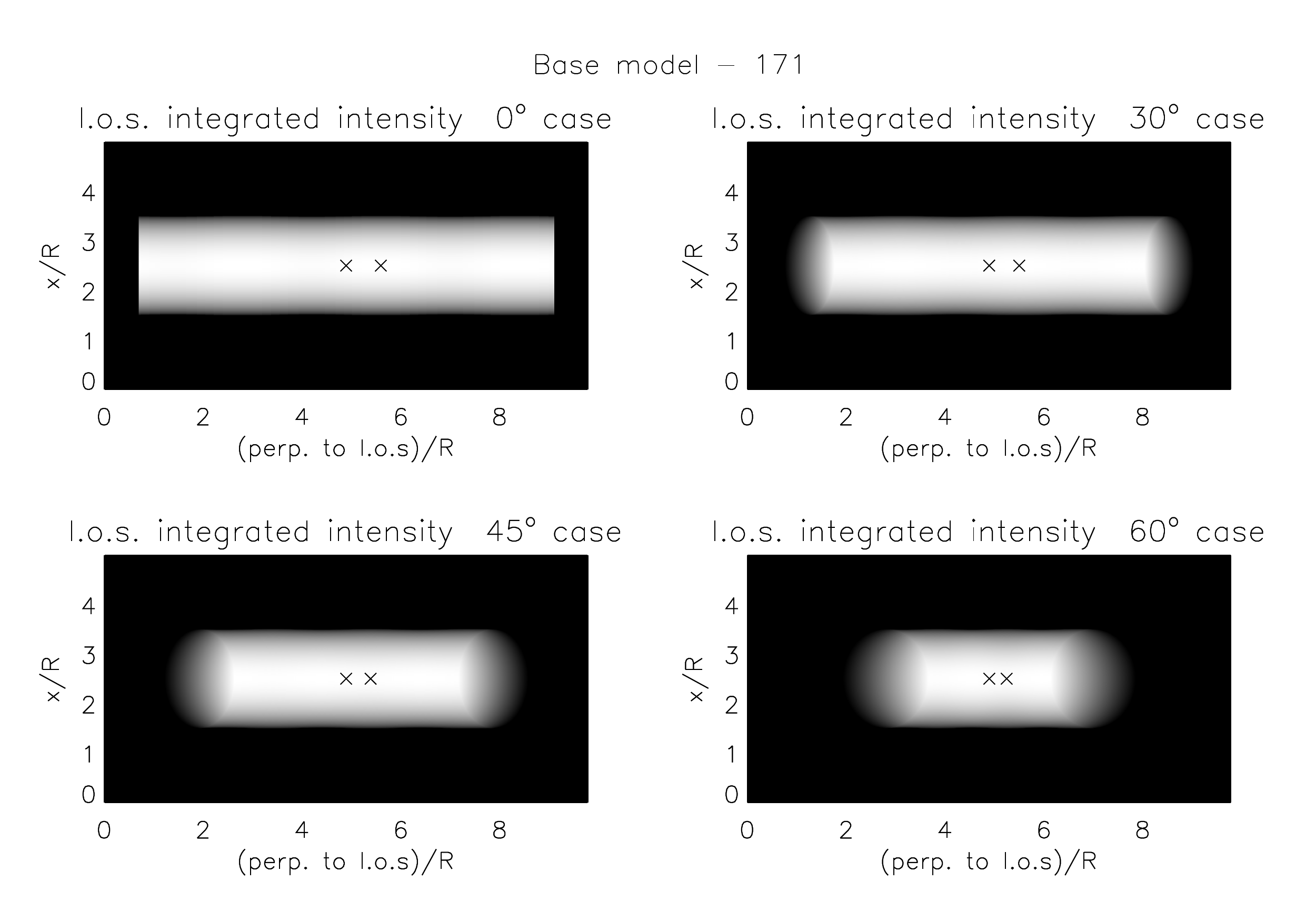}
\caption{Integrated intensity along the line-of-sight for the \textit{base model - 171}, at the 4 different angles considered: $0^{\circ}, 30^{\circ}, 45^{\circ}$ and $60^{\circ}$ (see Sect.~\ref{models} for more details). The snapshot is taken at 1/3 of the mode period. The crosses in each panel denote the location of the 2 line-of-sight rays defined in Sect.~\ref{spectrum}. In each panel, the left and right crosses correspond to the rays crossing a node and an anti-node of the oscillation.}
\label{fig3}
\end{figure}

\begin{figure}
\centering
\includegraphics[width=9cm]{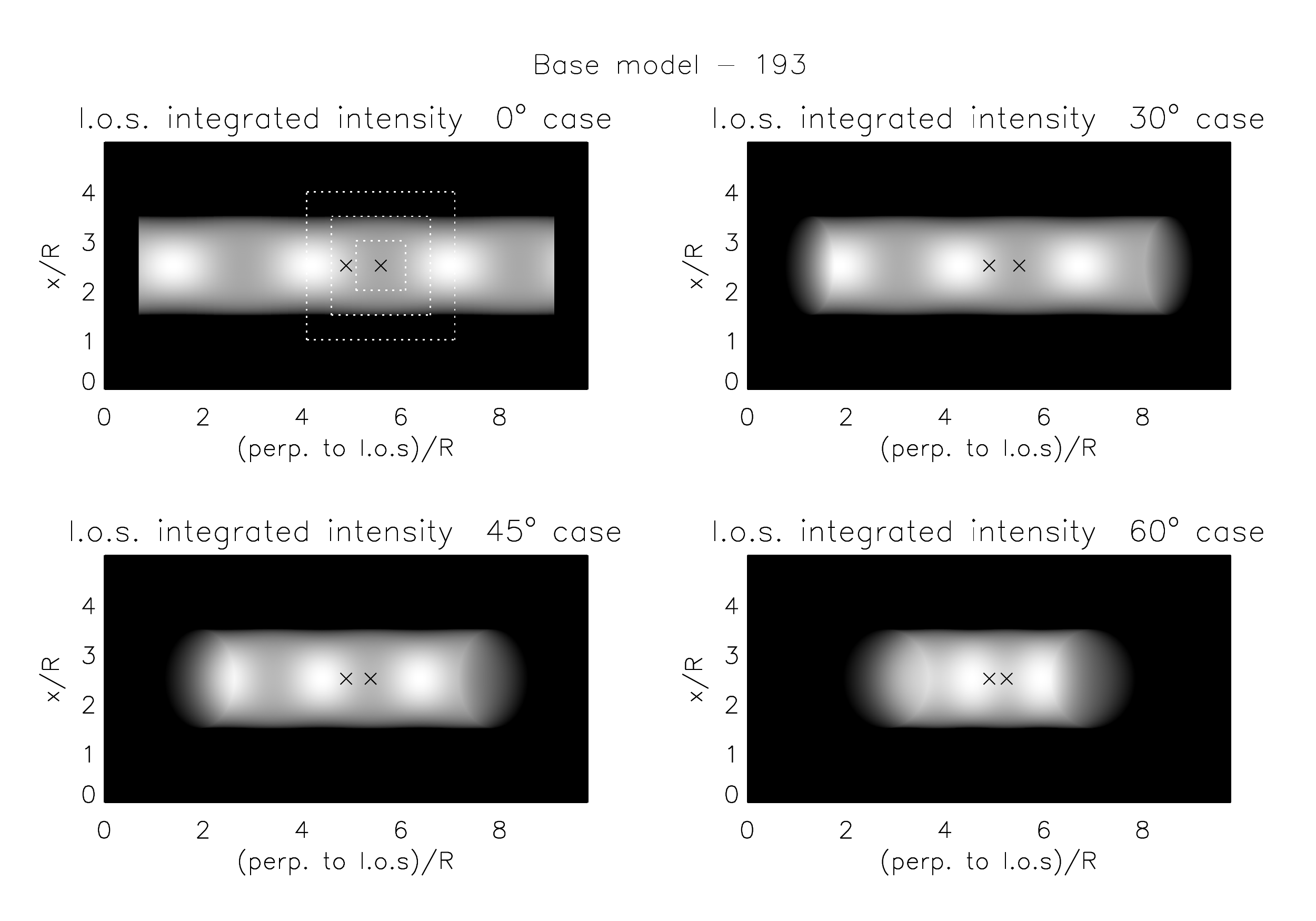}
\caption{Same as Fig.~\ref{fig3} but for the \textit{base model - 193} (see Sect.~\ref{models} for more details). The pixel sizes corresponding to 1R, 2R and 3R are shown around the location of ray 2 (anti-node).}
\label{fig4}
\end{figure}

\begin{figure*}[!htb]
\begin{center}$
\begin{array}{cc}
\includegraphics[width=6cm]{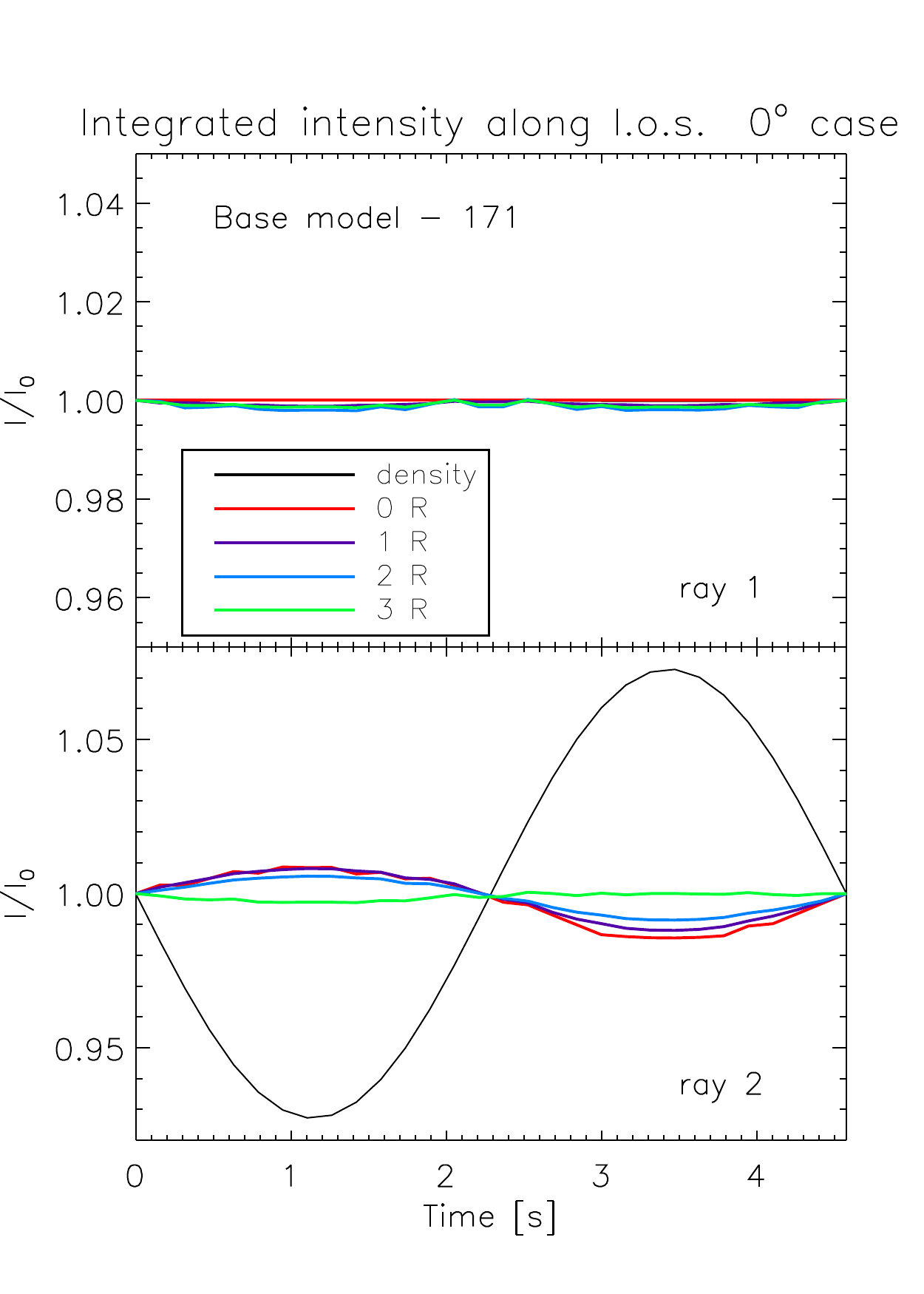} &
\includegraphics[width=6cm]{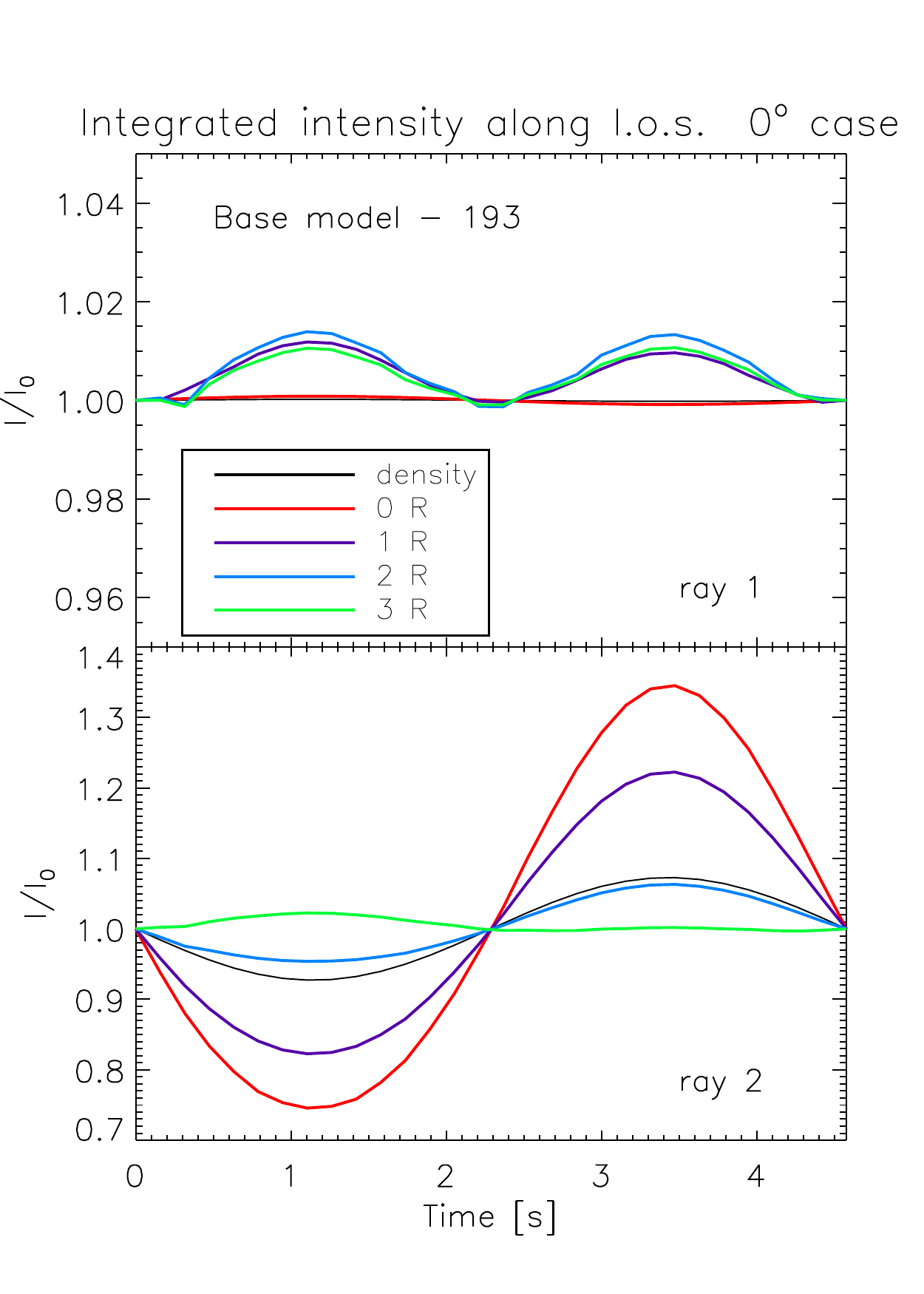}
\end{array}$
\end{center}
\caption{Integrated intensity along the line-of-sight (normalised with respect to the equilibrium value) for the $0^{\circ}$ angle case for the \textit{base model - 171} (left panels) and the \textit{base model - 193} (right panels) for two specific rays: ray 1 crossing a node (top most panels) and ray 2 crossing an anti-node (bottom panels), shown with crosses in Figs.~\ref{fig3} and \ref{fig4}. Different colors denote different pixel sizes: $0R, 1R, 2R$ and $3R$ in red, purple, blue and green, respectively, where $R=1~\mathrm{Mm}\approx\ell$ denotes the radius of the cylinder and $0R$ denotes the spatial resolution of our model: 25 km. The black curve denotes the density profile at the centre of the cylinder and the same location along the axis of the cylinder as rays 1 and 2. The intensity range of the top panels is set significantly smaller than that of the bottom panels in order to show the low intensity modulation at the nodes.}
\label{fig5}
\end{figure*}

\subsection{Models}\label{models}
The modulation of the intensity due to the sausage mode is analysed in several models which differ in 1 specific way from our base model. 
\begin{itemize}
\item[$\bullet$] \textbf{The `\textit{base model - 171}'} -- We analyze the modulation in intensity for our base model in the transition region line \ion{Fe}{ix} $171.073~\AA$, which has a formation temperature at $\log T=5.9~\mathrm{K}$. 
\item[$\bullet$] \textbf{The `\textit{base model - 193}'} -- We analyze the modulation in intensity for our base model in the coronal line \ion{Fe}{xii} $193.509~\AA$, which has a formation temperature at $\log T=6.1~\mathrm{K}$.
\item[$\bullet$] \textbf{The `\textit{high-T model}'} -- The effects of high external temperature are analysed in this model. We use an external temperature of $8\times10^{5}~\mathrm{K}$, which results in a slightly higher internal density, with a mean value at the anti-node of $5.5\times10^{-12}~\mathrm{kg~m}^{-3}$ and a variation around this value of roughly 7\%.
\item[$\bullet$] \textbf{The `\textit{long $\lambda$ model}'} -- Keeping the same values of internal and external thermodynamic quantities for the unperturbed cylinder (base model), here we model the longest possible wavelength for the sausage mode for these conditions: $\approx5~\mathrm{Mm}$, which results in a wavenumber of $kR=1.25$. Figure~\ref{fig2} shows a snapshot of a cut along the middle axis of the cylinder for the radial and longitudinal velocities, density and temperature for this case. The snapshot is taken at 1/3 of the period, which in this case is 5.8~s. The white and black lines show the line-of-sight rays at the different angles considered in this study. For this model two wavelengths in the $z$ direction are enough to cover the line-of-sight rays, giving $N_{z}=408$ grid points. The density and temperature have roughly the same average values as before but now exhibit slightly larger variations at the anti-nodes, about 8.5\% and 6\% respectively. The radial and longitudinal velocities have now maxima at $23~\mathrm{km~s}^{-1}$ and $3~\mathrm{km~s}^{-1}$.
\end{itemize}

\section{Results}\label{results}
Since the aim of this work is to identify the observational signatures of the sausage mode in coronal structures we present the results of our models as they would be perceived by the instruments at hand. We therefore divide the results into 2 relevant categories depending on the instrument: imaging and spectroscopic. 

\subsection{For imaging instruments}\label{imaging}

Figure~\ref{fig3} shows the integrated intensity for the \textit{base model - 171} (corresponding to the 10 times higher spatial resolution model), obtained by integrating the intensity in Eq.~\ref{intens} over the full \ion{Fe}{ix} $171~\AA$ line, for the 4 line-of-sight angles: $0^{\circ}, 30^{\circ}, 45^{\circ}$ and $60^{\circ}$. The crosses in the centre of each panel show the location of rays 1 and 2 crossing the cylinder defined in Sect.~\ref{spectrum}. The snapshot is taken at 1/3 of the period, coinciding with that of Fig.~\ref{fig1}. 

\begin{figure}
\centering
\includegraphics[width=8cm]{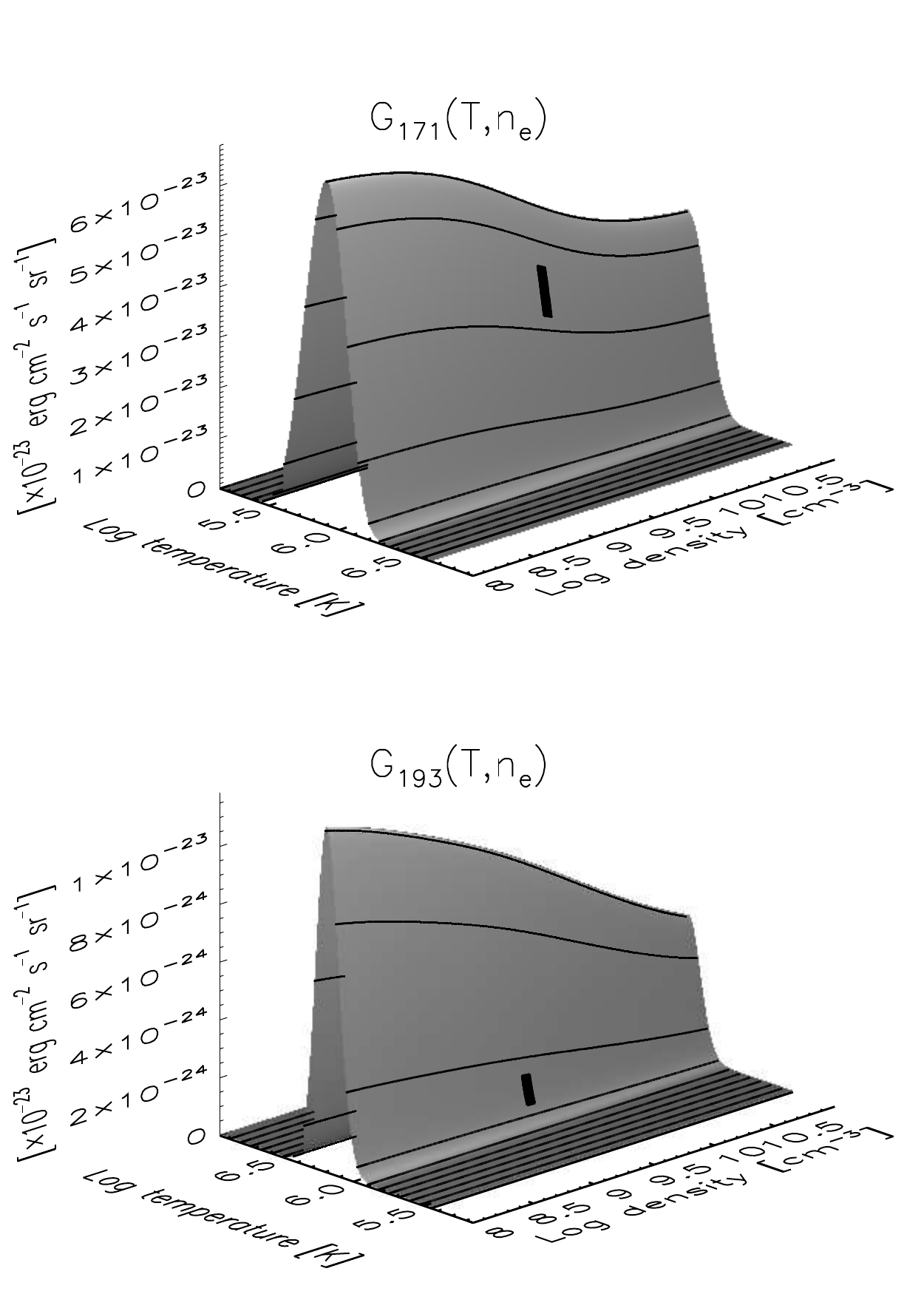}
\caption{Contribution function $G_{\lambda_0}(T,n_e)$ for \ion{Fe}{ix}~171 and \ion{Fe}{xii}~193. The maximum  formation temperature for these ions are, respectively, $\log \mathrm{T}=5.9$ and $\log \mathrm{T}=6.2$~K. The black parallelogram in each surface has its corners defined by the minimum and maximum temperature and density values found in the base model, and therefore includes the $G_{\lambda_0}(T,n_e)$ values used here. Note that in order to show the parallelogram for the $G_{171}(T,n_e)$ surface the temperature range has been inverted.}
\label{fig6}
\end{figure}

One of the main features of this figure is the absence of significant intensity structuring along the cylinder, in sharp contrast with the density or temperature maps shown in Fig.~\ref{fig1}. The intensity pattern matches the radial expansion of the cylinder, which is barely discernible, equal to only 1 pixel in the low resolution model (25~km). We thus have a longitudinal structuring in intensity qualitatively similar to the longitudinal structure of the radial velocity. On the other hand, contrary to what would be expected if the intensity is assumed to be proportional to the density squared, the regions of increased intensity correspond to regions of low density at the centre of the cylinder. This is due to the fact that the \ion{Fe}{ix} $171~\AA$ line has a formation temperature at $\log \mathrm{T}=5.9$, and the sausage mode is only slightly compressible in the corona, producing temperatures in our cylinder that oscillate between $\log \mathrm{T}=5.99$ and $\log \mathrm{T}=6.04$. Therefore, a drop in temperature in our model leads to an increase in the fraction of the \ion{Fe}{ix} population, thus increasing the emission, as can be seen in Figure~\ref{fig6}. This effect is further reinforced by the fact that at $0^{\circ}$ viewing angle the rays crossing these regions will cross more material since they correspond only to regions of expansion (lower temperature). On the other hand, as shown in Figure~\ref{fig6}, the \ion{Fe}{xii}~$193~\AA$ line has a formation temperature at $\log \mathrm{T}=6.2$~K. In this case, higher temperature implies an increase in the ion population, leading to higher intensity and thus creating a structure that is similar to the temperature or density structure, as can be seen in the \textit{base model - 193} shown by Figure~\ref{fig4}. This fact highlights the dominant role of the contribution function $G_{\lambda_0}(T,n_e)$ in the formation of the emergent intensity. It is important to remind the reader that these results are subject to the assumption of equilibrium ionisation, and therefore that the plasma responds instantaneously to the changes in temperature and density from the MHD mode. The limitations of this assumption will be discussed at the end of this section.


Figure~\ref{fig5} shows the evolution of the integrated intensity along the line-of-sight at an angle of $0^{\circ}$ for the \textit{base model - 171} (left panels) and the \textit{base model - 193} (right panels), for the rays 1 (top panels) and 2 (bottom panels) crossing at a node and an anti-node of the oscillation, respectively. Any other location along the axis of the cylinder displays features similar to either ray 1 or ray 2. For each ray different spatial resolution pixels are taken, marked in different colors (see figure for details). In the case of the \textit{base model - 171} we have a very low modulation of the intensity, as compared to the density modulation at the centre of the cylinder (black curve in the figure). While the latter varies by $\lesssim7\%$, the intensity varies by $\lesssim2\%$ at an anti-node and is basically negligible at the location of a node. Furthermore, as explained earlier, the variation of the intensity is in antiphase with that of the density due to the fact that the contribution function $G_{171}(T,n_e)$ has a peak at lower temperature values than those of the plasma in the cylinder. The slightly longer column thickness created by the area expansion  creates the slight asymmetry between the two dips of the intensity cycle.

This scenario is in sharp contrast with that of the \textit{base model - 193}, shown in the right panels of Fig.~\ref{fig5}. The $G_{193}(T,n_e)$ function presents a peak at a higher temperature value than that of the plasma inside the cylinder. In this case we thus have in-phase intensity variations with the density (or temperature) profile. Also, as shown in Figure~\ref{fig6}, the $G_{193}(T,n_e)$ contribution function is significantly more sensitive to the temperature changes inside the cylinder than the $G_{171}(T,n_e)$ function, leading to intensity variations of $\simeq35~\%$. The difference in intensity from the initial value is greater for the maximum values than the minimum values. This is due to the fact that the gradient of the $G_{\lambda_{0}}(T,n_e)$ function is steeper for both 171 and 193 above its initial value (mainly in the direction of higher temperature) than below the initial value. This effect is further amplified by the density enhancement at a time of contraction, since each pixel along the line-of-sight contributes with the squared of the local density value. We further notice 2 small bumps for ray 1 at times of density extrema when considering pixel sizes different from 0R. These are again due to the combined effect of the steepness of the $G_{\lambda_{0}}(T,n_e)$ gradient and enhanced density in regions of contraction, which are within the line-of-sight for coarse pixel sizes.

\begin{figure}[!h]
\centering
\includegraphics[width=6cm]{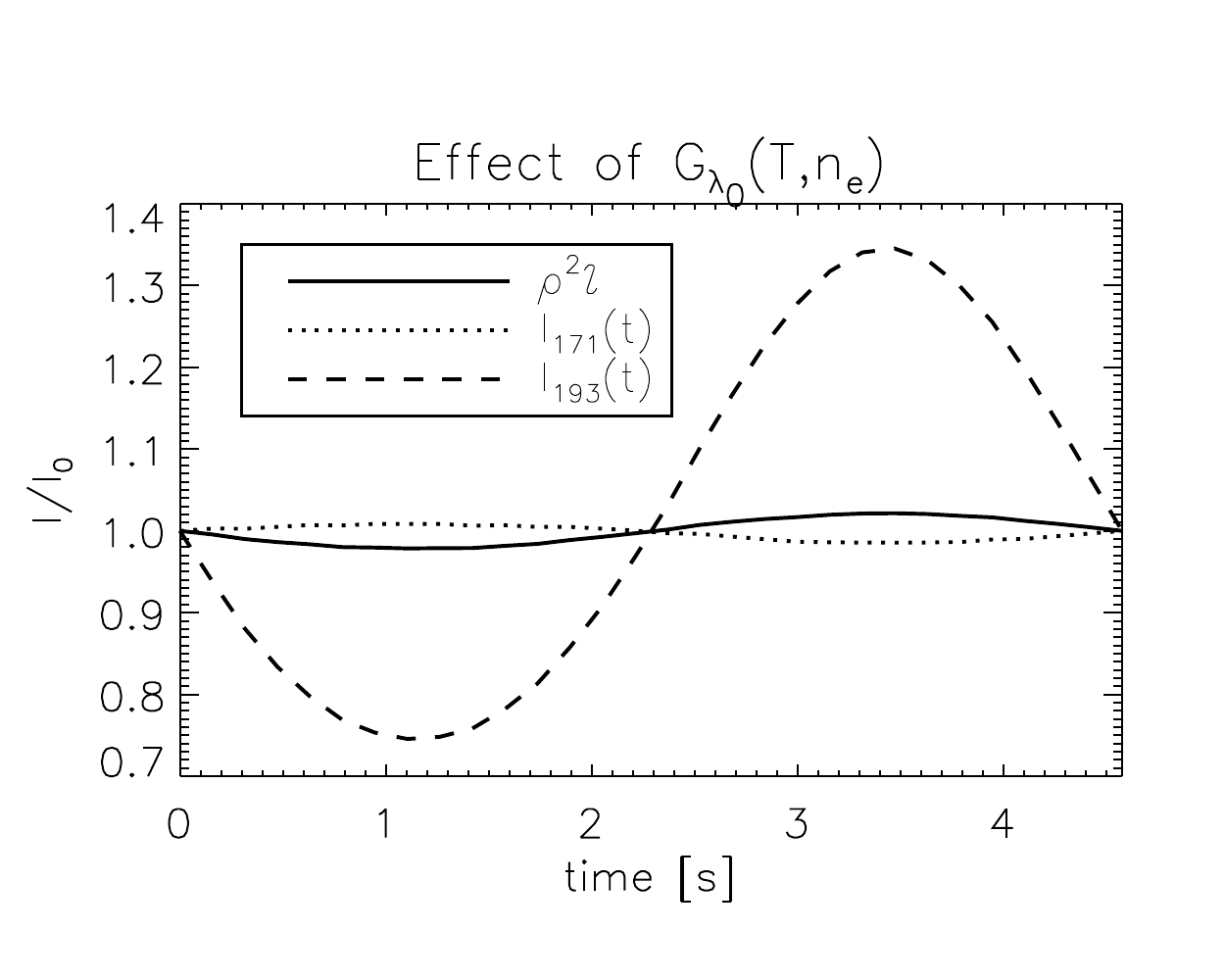}
\caption{Comparison between an intensity calculated as $I\approx\rho^2 l$, where $l$ is the column thickness along the line-of-sight (solid), the \textit{base model - 171} intensity $I_{171}(t)$ (dotted) and the \textit{base model - 193} intensity $I_{193}(t)$ (dashed) for a viewing angle of $0^{\circ}$ and a pixel size resolution of 0R for ray 2. The differences between the lines illustrates the importance of the $G_{\lambda_0}(T,n_e)$ function.}
\label{fig7}
\end{figure}
\begin{figure}[!h]
\centering
\includegraphics[width=6cm]{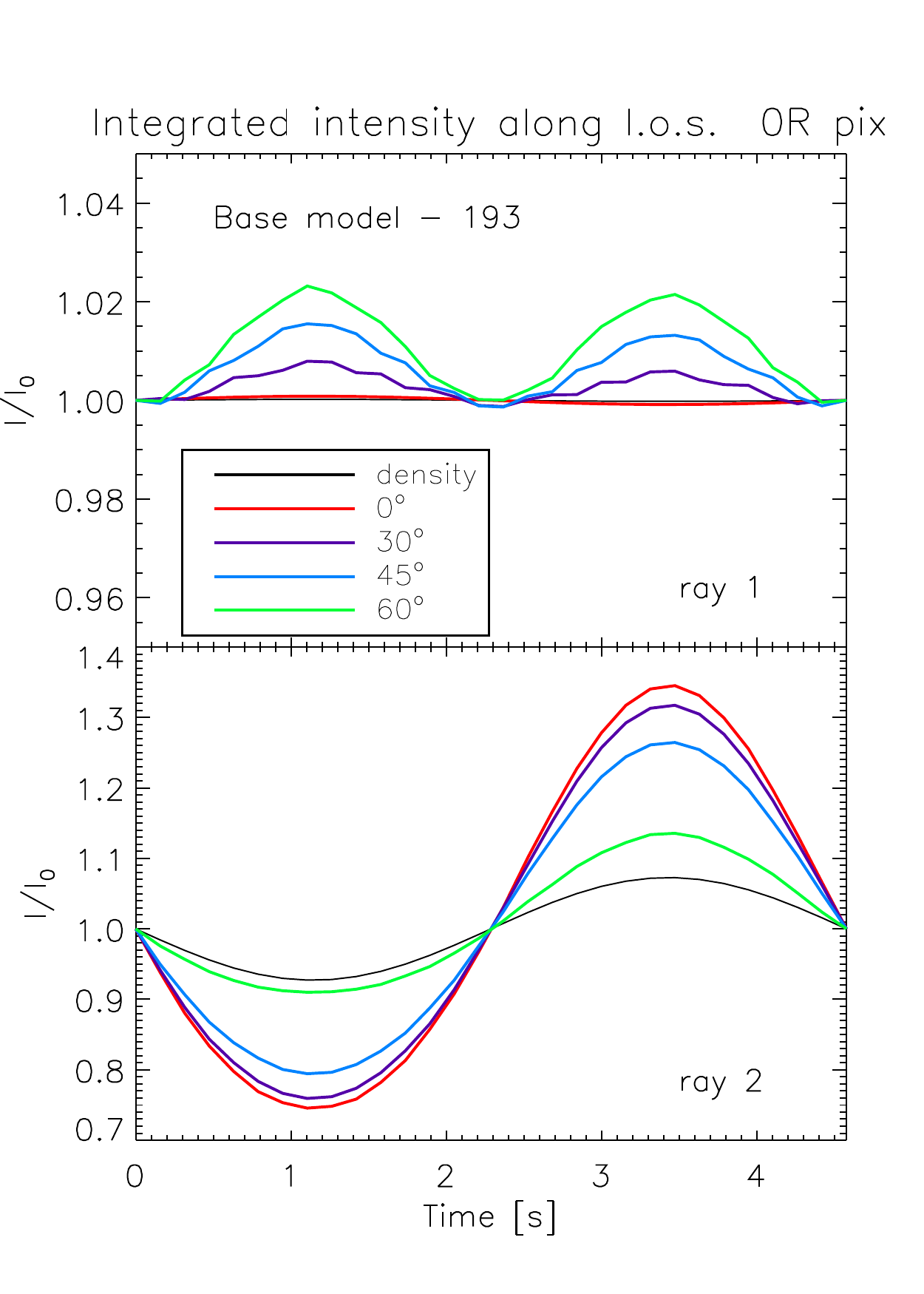}
\caption{Same as the right panel of Fig.~\ref{fig5} but fixing the pixel size resolution to 0R and varying the angle of the line-of-sight for the \textit{base model - 193}. $0^{\circ}, 30^{\circ}, 45^{\circ}$ and $60^{\circ}$ correspond to the red, purple, blue and green curves, respectively.}
\label{fig8}
\end{figure}

The variation of the intensity decreases significantly with the pixel size for all models, especially at anti-node locations when viewing at a $0^{\circ}$ line-of-sight angle. In the considered models the intensity variation is reduced by factors of $10-17$ when increasing the pixel size to 3R. This is expected since a larger pixel size implies a larger probability of crossing mixed regions of expansion and contraction. For the models with a wavenumber $ka=2.24$, the unit of the pixel is slightly larger than $\frac{1}{3}$ of the wavelength, and therefore 3 pixel units ($\simeq3$R) not only covers the entire region of expansion and contraction, but also produces a slight imbalance between the two, as can be seen in Figs.~\ref{fig3} and \ref{fig4}. Hence, at a time of contraction the pixel will cover slightly more expanding material, and vice-versa, resulting in a reversal of the variation in intensity at anti-node locations (ray 2). This can be seen in Fig.~\ref{fig5} for both models, 171 and 193. However, in this case, most of the longitudinal intensity structuring is erased, and with it the difference between the 171 and the 193 line intensity variation. Furthermore, if our pixel is centred in a node, regardless of the size of the pixel the amount of expansion and contraction will always be similar, thus producing minimal intensity variation. As can be seen in the top right panel of Fig.~\ref{fig5}, this variation attains a maximum for a 2R pixel size, which is linked to the fact that the amount of material entering and leaving the line-of-sight is close to a maximum for this size. This is of course dependent on the shape of the pixel, which in our case is a square.

\begin{figure}[!h]
\begin{center}$
\begin{array}{c}
\includegraphics[width=9cm]{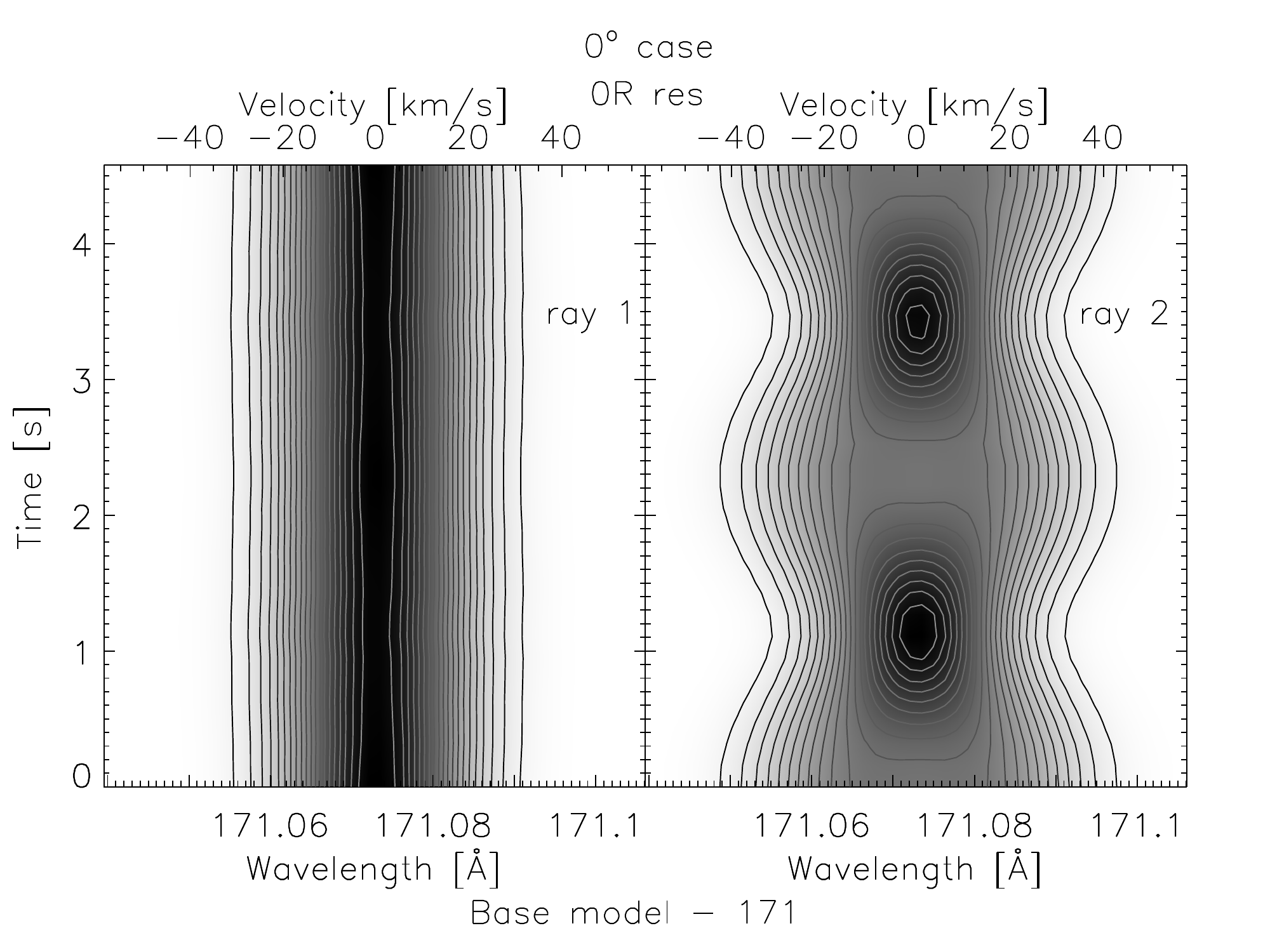} \\
\includegraphics[width=9cm]{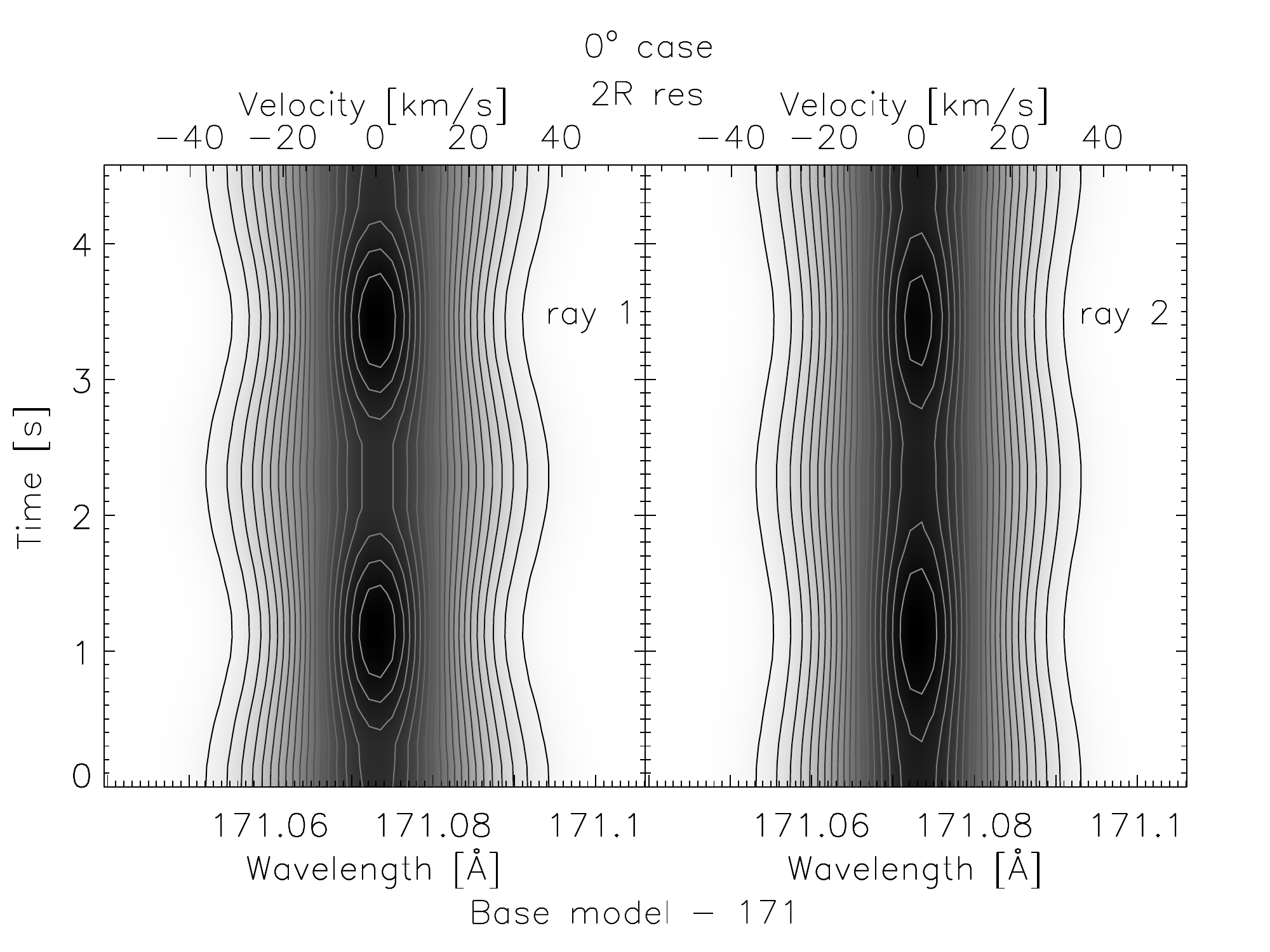}
\end{array}$
\end{center}
\caption{Time variation of the specific intensity $I_{\lambda}$ integrated along ray 1 (left panels) and ray 2 (right panels) for the \textit{base model - 171} when viewing at an angle of $0^{\circ}$ and for a pixel size of 0R (upper panels) and 2R (lower panels). Contours of the intensity are also shown for better visualization of the variation.}
\label{fig9}
\end{figure}
\begin{figure}[!h]
\begin{center}$
\begin{array}{c}
\includegraphics[width=9cm]{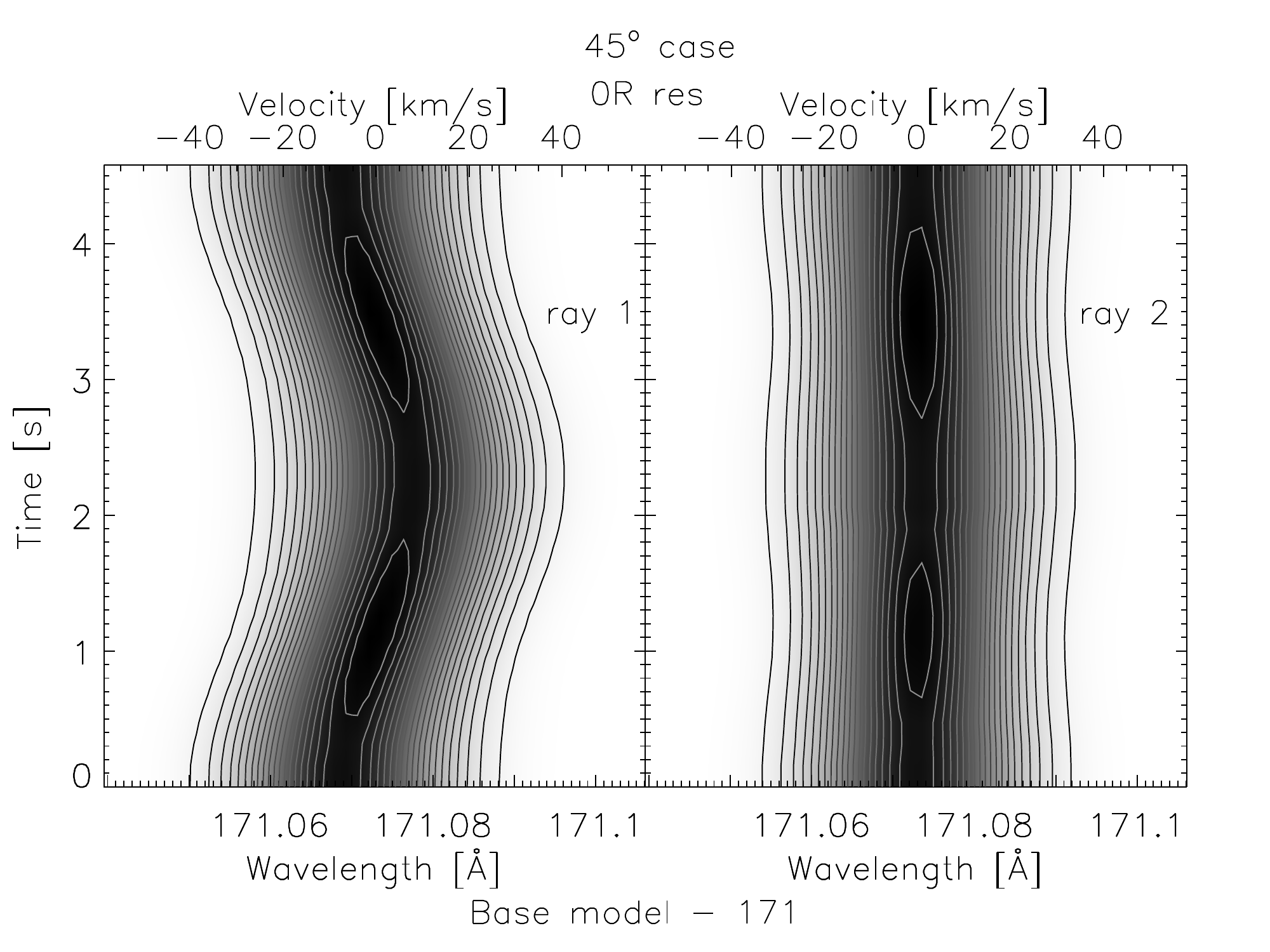} \\
\includegraphics[width=9cm]{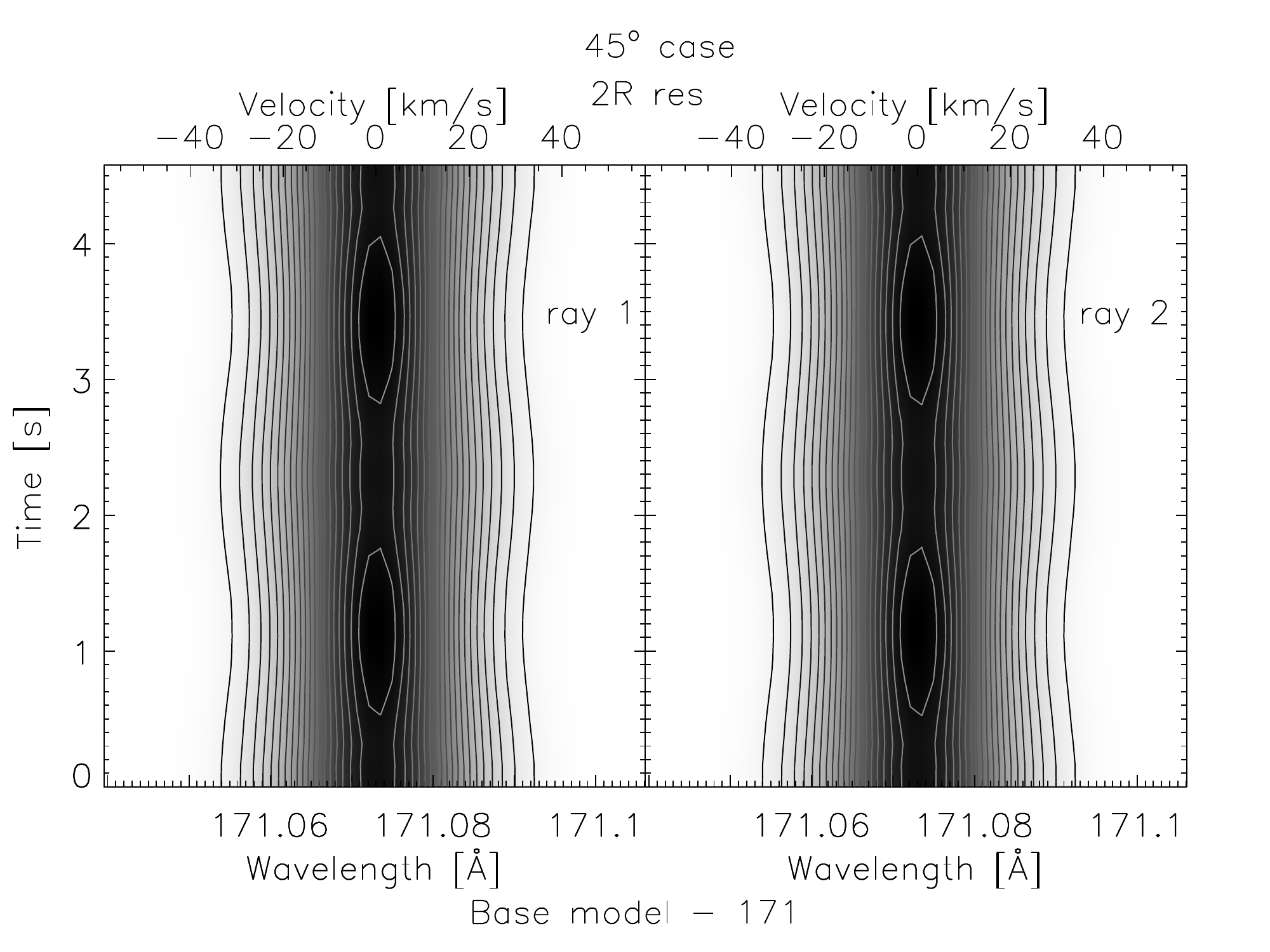}
\end{array}$
\end{center}
\caption{Same as Fig.~\ref{fig9} but for a line-of-sight crossing the cylinder at an angle of $45^{\circ}$.}
\label{fig10}
\end{figure}

As explained in the introduction, a common approximation to intensity calculation is $I\approx\rho^2 l$, where $l$ is the column thickness along the line-of-sight. In Fig.~\ref{fig7} we compare this approach to the outcomes of the more proper intensity calculation performed in this work, plotting $\rho^2 l$ and the results from the \textit{base model - 171} and \textit{base model - 193}. To clearly show the difference we plot the case of a $0^{\circ}$ viewing angle and a pixel size resolution of 0R for ray 2 (anti-node). By following the $I\approx\rho^2 l$ approximation we find maximum intensity variations below 3~\%, which match those obtained with the 171 line. However, both profiles are in antiphase with each other due to the fact that the \ion{Fe}{ix} ion population decreases with increasing temperature for the range of temperatures considered here. The discrepancy in amplitude with the 193 line intensity is obvious. This illustrates the important role of the contribution function $G_{\lambda_{0}}(T,n_e)$ in the formation of the emergent intensity. 

The variation in the integrated intensity due to different line-of-sight angles is shown in Fig.~\ref{fig8}. In order to show best the variation we fix the pixel size resolution at 0R and consider the \textit{base model - 193}, for which large intensity variation is found for the temperature and density variations created by the sausage mode. By increasing the line-of-sight angle the rays cross more regions of hot (contraction) and cold (expansion) plasma, thus reducing the variation of the intensity by a factor of 2.5 (down to $\sim10~\%$). For rays crossing nodes (ray 1), as discussed previously, the variation is greatly reduced from the  crossing of mixed temperature regions in equal quantities along the rays. As in Fig.~\ref{fig5} (right panel), two small bumps can be seen for non-zero viewing angles produced by the combined effect of the local gradient steepness of the $G_{\lambda_0}(T,n_e)$ function and the regions of enhanced density along the line-of-sights.

Apart from the 2 models discussed in this section, the other models exhibit the same behaviour as that of the \textit{base model - 171}, with minor quantitative differences. The main differences are discussed in section~\ref{diffmodel}.

\begin{figure*}[!ht]
\begin{center}$
\begin{array}{cc}
\includegraphics[width=6cm]{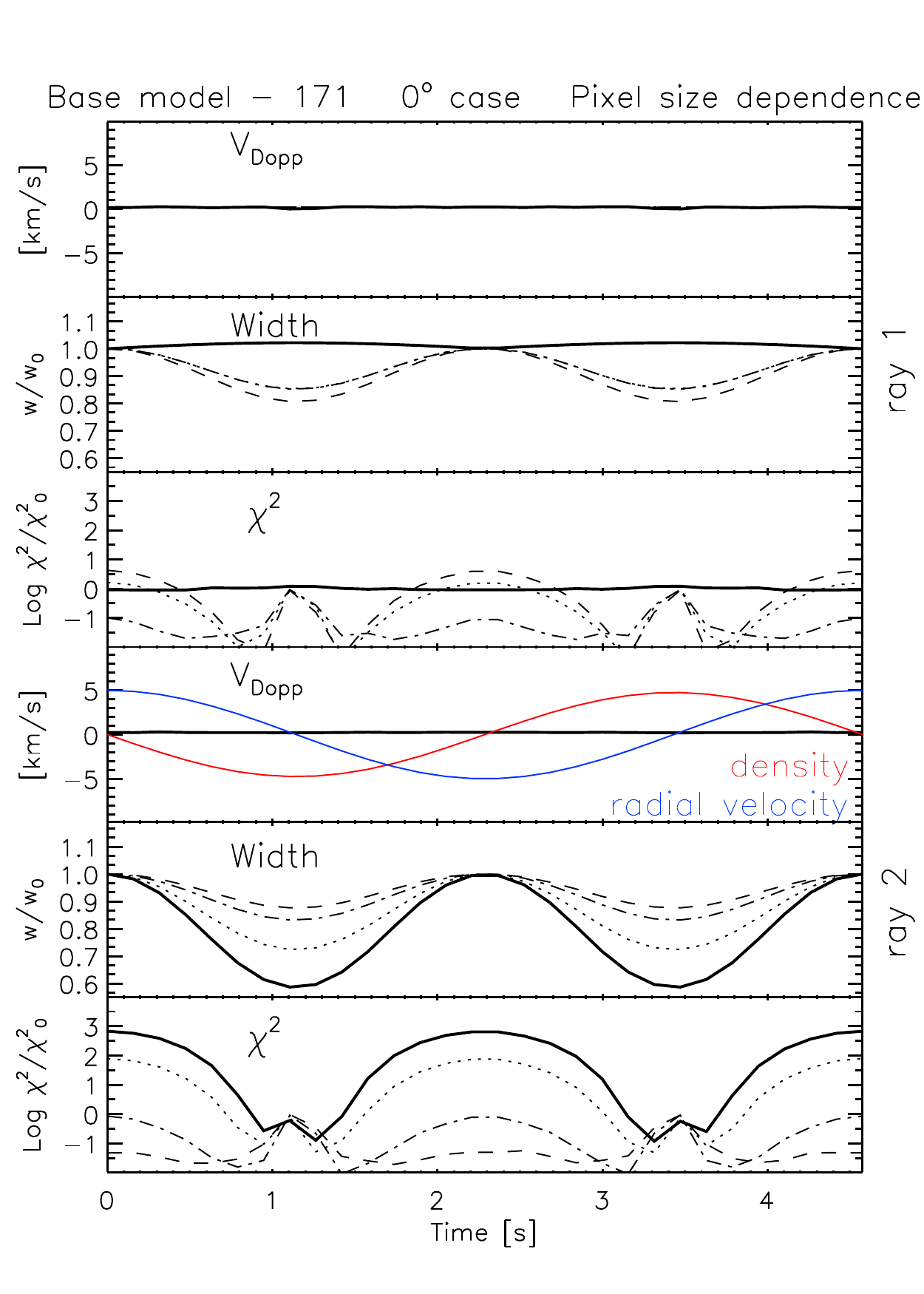} &
\includegraphics[width=6cm]{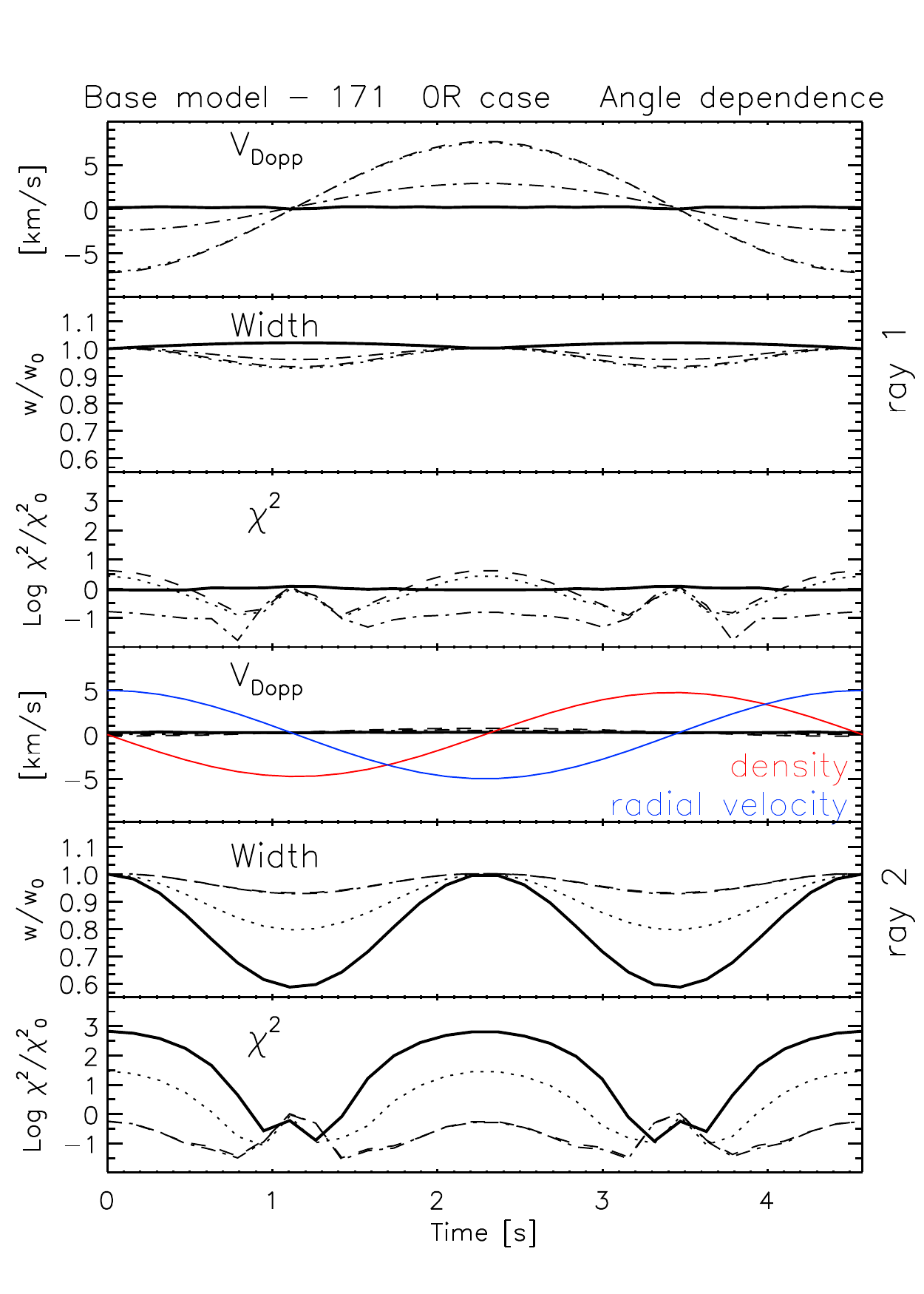}
\end{array}$
\end{center}
\caption{\textit{Left panel:} Pixel size resolution dependence of the time variation of the specific intensity $I_{\lambda}$ for the $0^{\circ}$ angle case of the \textit{base model - 171}. From top to bottom we show quantities ensuing from a Gaussian fit to the spectral line. The first 3 upper panels correspond to the Doppler velocity, the line width and the goodness-of-fit $\chi^{2}$ parameter for the case of ray 1 (node). The 3 lower panels show the same quantities for the case of ray 2 (anti-node). Different line styles correspond to different pixel size resolution: 0R (solid), 1R (dotted), 2R (dashed), 3R (dot-dashed). The line width is normalised with respect to the initial value at time $t=0$. The unit for the $\chi^{2}$ parameter is the logarithm of the normalised value, where the normalisation factor is the equilibrium value. \textit{Right panel:} Angle dependence of the time variation of the specific intensity $I_{\lambda}$ for the 0R pixel size case of the \textit{base model - 171}. The arrangement of these panels is the same as for the left panels. Different line styles denote different line-of-sight angles: $0^{\circ}$ (solid), $30^{\circ}$ (dotted), $45^{\circ}$ (dashed), $60^{\circ}$ (dot-dashed). For both left and right panels the variation in time of density (red curve) and radial velocity (blue curve) at the centre of the tube is plotted for reference in arbitrary units in the Doppler velocity panel.}
\label{fig11}
\end{figure*}

\subsubsection*{Non-equilibrium ionisation effects}

Perhaps the most important assumption in the intensity modulation presented here is to consider that the plasma is at all times in equilibrium ionisation, i.e. that the plasma response to the temperature and density changes produced by the MHD mode is instantaneous. However, this is never the case and significant departures from equilibrium can occur in timescales which can be short with respect to the timescale of the driver \citep{DeMoortel_Bradshaw_2008SoPh..252..101D}. The population densities of the ions in the plasma will vary according to the ionisation and recombination rates, which depend on the thermodynamic conditions at each specific moment of time. The rigorous solution to this problem involves solving the time-dependent ion balance equations, which would take us outside the scope of this paper. However, we can estimate the consequences of non-equilibrium ionisation on the emergent intensity. For this we assume that the plasma follows a Maxwellian distribution at all times. Timescales for ionisation and recombination equilibrium for the ions \ion{Fe}{ix} and \ion{Fe}{XII} can then be obtained through CHIANTI. These values are shown in Table~\ref{table2} \citep[in the case of a ${\kappa}$-distribution, as shown by][, the rates would increase leading to smaller timescales than those presented here]{Dzifcakova_2002SoPh..208...91D}.

For \ion{Fe}{ix} we can see that the ionisation and recombination timescales to reach equilibrium are on the order of 2~s, which is almost half the period of the sausage mode. This means that at a time of expansion (cooling) the material will just have started being ionised and will therefore start recombining without attaining a full ionisation. Hence, we have a time lag effect and an attenuation (and deformation) of the intensity profile. The loss in amplitude will be similar to that obtained when integrating the profile over a period of 2~s, in which case the intensity is reduced by a factor of roughly 1.5. This effect is shown in Figure~\ref{fig13}. For \ion{Fe}{xii} the ionisation and recombination timescales are between 13~s and 20~s, that is, considerably larger than the period of our mode. In this case, there would be basically no modulation of intensity.

Non-equilibrium ionisation effects would however have no influence on the spectrometric results presented in section~\ref{spectro}. Indeed, as we will see in the next section, most of the spectral broadening is of turbulent nature and not of thermal origin.

\begin{table*}
\caption{\label{table2} Estimates of timescales for ionisation (I) and recombination (R) equilibrium for \ion{Fe}{ix} and \ion{Fe}{xii} inside the model}
\centering
\begin{tabular}{c c c}
\hline
\hline
\multirow{2}{*}{Ion} & $T_{min}\approx1\times10^6~\mbox{\rm{K}}$ & $T_{max}\approx1.1\times10^6~\mbox{\rm{K}}$ \\
 & $n_{e,~min}\approx2.7\times10^9~\mbox{\rm{cm}}^{-3}$ & $n_{e,~max}\approx3\times10^9~\mbox{\rm{cm}}^{-3}$ \\
\hline
\ion{Fe}{ix} & 2~s (I), 2.1~s (R) & 1.3~s (I), 2.1~s (R) \\
\ion{Fe}{xii} & 22.5~s (I), 1.5~s (R) & 12.9~s (I), 1.5~s (R) \\
\hline
\end{tabular}
\tablefoot{The timescales are given in seconds and are calculated as $1/(n_e\times X(T))$, where $X(T)$ denotes either the ionisation or recombination rate obtained with CHIANTI assuming a Maxwellian distribution for the plasma. The temperature ($T$) and number density ($n_e$) values used for the calculation are extremum values found in the centre of the cylinder.}
\end{table*}

\subsection{For spectroscopic instruments}\label{spectro}

\begin{figure}
\centering
\includegraphics[width=6cm]{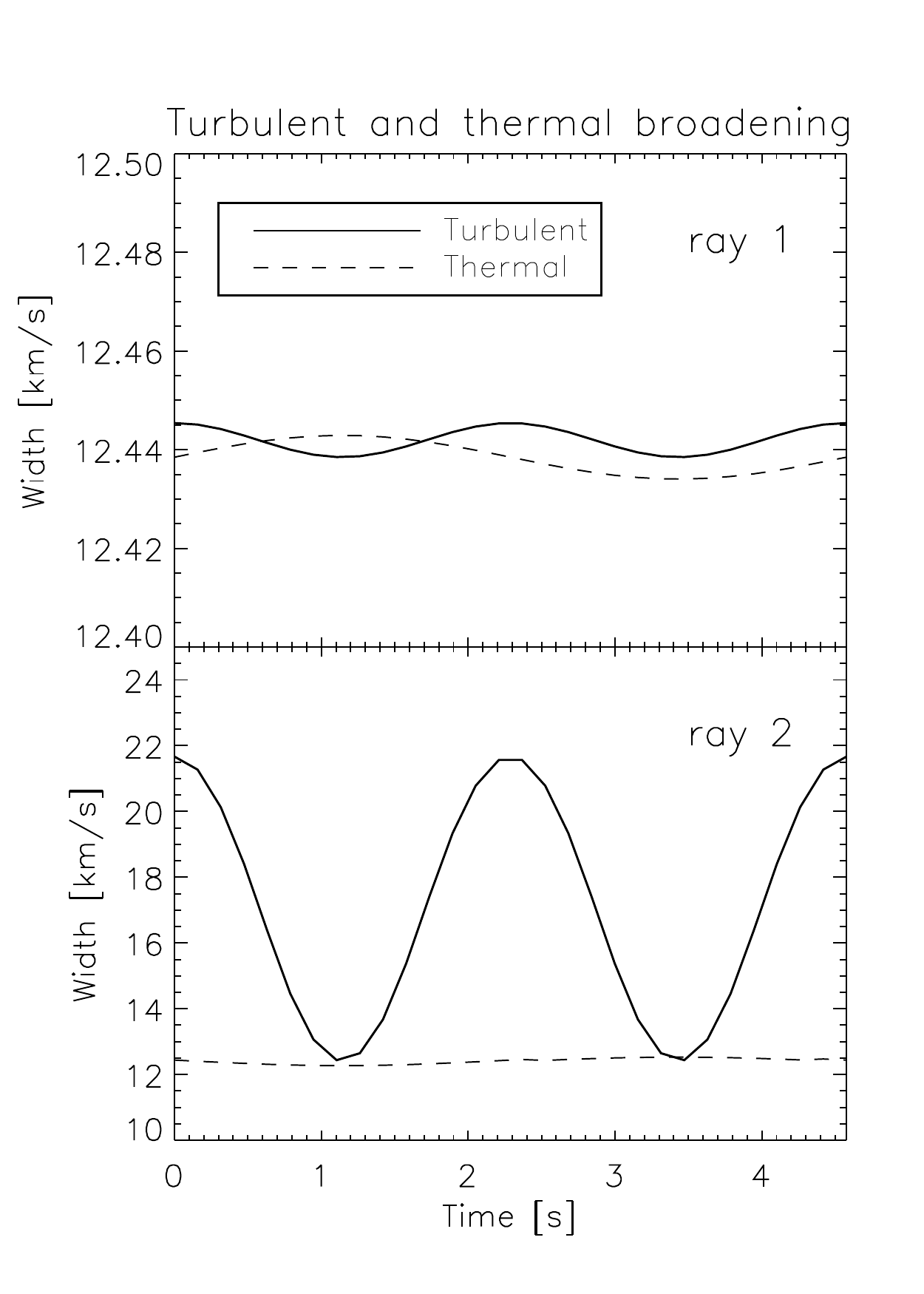}
\caption{Estimation of the spectral line broadening contribution from the unresolved velocity perturbations (which we denote as turbulence) and temperature perturbations. The Gaussian width corresponding to the thermal broadening contribution (dashed curve) is obtained from the average of temperature for the pixels along the line-of-sight crossing the cylinder. The Gaussian width corresponding to the turbulent broadening contribution (solid lines) is obtained from a Gaussian fit to the sum of multiple Gaussians, one for each pixel in the cylinder along the line-of-sight, with a fixed temporal averaged thermal broadening and shifted centre given by the line-of-sight velocity at that pixel.}
\label{fig12}
\end{figure}

We now analyze the spectral variation caused by the sausage mode, which is of interest for observations performed by spectroscopic instruments. For this we concentrate on rays 1 and 2 only, simulating slits that would cross the cylinder at a perpendicular angle, centred at a node or an anti-node, respectively. All other positions along the cylinder present features similar to either ray 1 or ray 2. 

Figures~\ref{fig9} and \ref{fig10} show the time variation of the specific intensity $I_{\lambda}$ for the \textit{base model - 171} for the line-of-sight angles of $0^{\circ}$ and $45^{\circ}$ respectively. Upper and lower panels in the figures denote, respectively, 0R and 2R pixel size resolution, and left and right panels denote, respectively, ray 1 (node) and ray 2 (anti-node). For the $0^{\circ}$ angle and 2R pixel size case we have periodic spectral broadening for rays crossing the cylinder at any location, especially clear for the 0R pixel size case of ray 2. This feature is only absent for the case of ray 1 with 0R pixel size. This is expected since for the latter the pixel size does not include any ray crossing a region of expansion or contraction, contrary to all the other cases. When watching at an angle, rays will cross regions of expansion and contraction by different amounts when the slit is centred at a node, and in equal amounts when centred at an anti-node. This creates an asymmetry for the former and therefore an overall dominating redshift or blueshift. This translates in clear periodic blueshift or redshift excursions for the 0R pixel size case of ray 1 (left upper panel of Fig.~\ref{fig10}), but hardly discernible for the 2R pixel size case (lower left panel). As expected, the periodic broadening is also present for this viewing angle for any pixel size.

The pixel size resolution dependence of the time variation of the specific intensity $I_{\lambda}$ observed in Figs.~\ref{fig9} and \ref{fig10} is shown in the left panels of Fig.~\ref{fig11} for the $0^{\circ}$ angle case. In these panels we show the result of Gaussian fits to the specific intensity $I_{\lambda}(t)$ for all the considered pixel sizes. The time variation of the Doppler velocity, width of the Gaussian fit, and the goodness-of-fit $\chi^{2}$ statistic are shown from top to bottom in the first 3 panels for ray 1 and the last 3 panels for ray 2. Since the $0^{\circ}$ angle case is considered and the sausage mode is mainly a radial mode, negligible Doppler velocity results from the Gaussian fit for any pixel size resolution. On the other hand, the line width and the non-Gaussianity $\chi^{2}$ parameter present significant variation, and have profiles that are in phase with the square of the radial velocity. This means that they both have a double periodicity. Maximum variations close to 20~\% and 50~\% are obtained for the line width, respectively for rays 1 and 2. Notice that for the case of ray 1 (ray 2) the spectral line broadening increases (decreases) for pixel sizes up to 2R. For 3R, which corresponds to 1 wavelength, and therefore an equal amount of expanding and contracting gas, the broadening is reduced (increased) for ray 1 (ray 2). The decreasing factors range between 2.5 and 3.4 between the models. The temperature (or density) modulation by the sausage mode contributes to a certain extent to the non-Gaussianity of the line profile. This is especially the case for large pixel sizes ($\gtrsim2$R), for which the contribution from the radial velocity is lowest due to the large range of velocities included, producing a broad Gaussian-like profile. This can be seen in the $\chi^{2}$ profiles as a bump at times of maximum and minimum densities at the centre of the cylinder. These bumps are of the same order as the perturbations created by the radial velocities, and therefore can give to an observer the impression of a quadruple periodicity for the $\chi^{2}$ parameter (period of $P/4$ where $P$ is the period of the sausage mode).

The angle dependence of the time variation of the specific intensity $I_{\lambda}$ for rays 1 and 2 is shown in the right panels of Fig.~\ref{fig11} for the 0R pixel size resolution case of the \textit{base model - 171}. For non-zero viewing angles the rays now cross both regions of expansion and contraction. The Doppler velocity values ensuing from the Gaussian fits are therefore the emission measure weighted averages over these regions. A small contribution comes also from the longitudinal velocity component, but it is however one order of magnitude smaller than the transverse velocity component. As observed in Fig.~\ref{fig10}, depending on the line-of-sight angle one can have at a given time either a dominating blueshift ($30^{\circ}$ and $45^{\circ}$) or redshift ($60^{\circ}$). The variations in spectral line broadening given by the line width and the $\chi^{2}$ parameter decrease significantly with the angle when crossing an anti-node (with maximum factors for line width between 5.6 and 11.3 between the models). On the other hand, rays crossing a node have slightly increasing variations of spectral line broadening with angle (below 10~\%). The maximum velocity component is directed perpendicularly to the cylinder axis. The components creating the width of the spectral line therefore decrease with the sine of the line-of-sight angle. The contribution to the non-Gaussianity of the line from the temperature and density profiles can also be observed at various angles. The bumps in the $\chi^{2}$ parameter profile created at times of maximum or minimum temperature (and density) at the centre of the cylinder are of the same order of the perturbation created by the radial motions for any angle in the case of ray 1 and for angles $\gtrsim45^{\circ}$ for ray 2. 

In order to estimate the respective contribution from turbulence (unresolved velocity perturbations) and temperature to the broadening, we plot in Figure~\ref{fig12} the time variation of 2 Gaussian widths obtained from rays 1 and 2 with a viewing angle of $0^{\circ}$ and and pixel size of 0R. The Gaussian width corresponding to the thermal broadening contribution (dashed curve) is obtained from the average of temperature for the pixels along the line-of-sight crossing the cylinder. The Gaussian width corresponding to the turbulent broadening contribution (solid lines) is obtained from a Gaussian fit to the sum of multiple Gaussians, one for each pixel in the cylinder along the line-of-sight, with a fixed temporal averaged thermal broadening and shifted centre given by the line-of-sight velocity at that pixel. We can see that the turbulent broadening is the main contributor to the spectral line broadening. Only for cases of little plasma motion along the line-of-sight ($0^{\circ}$ angle and 0R pixel size, or at times of 0 radial velocity) is the thermal broadening comparable to the turbulent broadening.

\subsection{Effects of temporal and wavelength resolutions}\label{cad_wav}

In section~\ref{imaging} we discussed the effects from non-ionisation equilibrium, and in the last section we showed the effects of spatial resolution and line-of-sight angle on the detection of intensity modulation by the sausage mode. In this section we would like to present the main effects due to temporal and wavelength resolutions. The temporal and wavelength resolutions of our base model are, respectively, 0.15~s and 1.4~m$\AA$. Here we will discuss the differences introduced by taking temporal resolutions of 1~s and 2~s (0.15~s, 1~s and 2~s correspond, respectively, to 3~\%, 22~\% and 44~\% of the mode period), and wavelength resolutions of 4.2~m$\AA$ and 8.5~m$\AA$ (1.4~m$\AA$, 4.2~m$\AA$ and 8.5~m$\AA$ correspond, respectively, to 48, 16 and 8 points sampling of the spectral line). 

\begin{figure}[!h]
\begin{center}
\includegraphics[width=6cm]{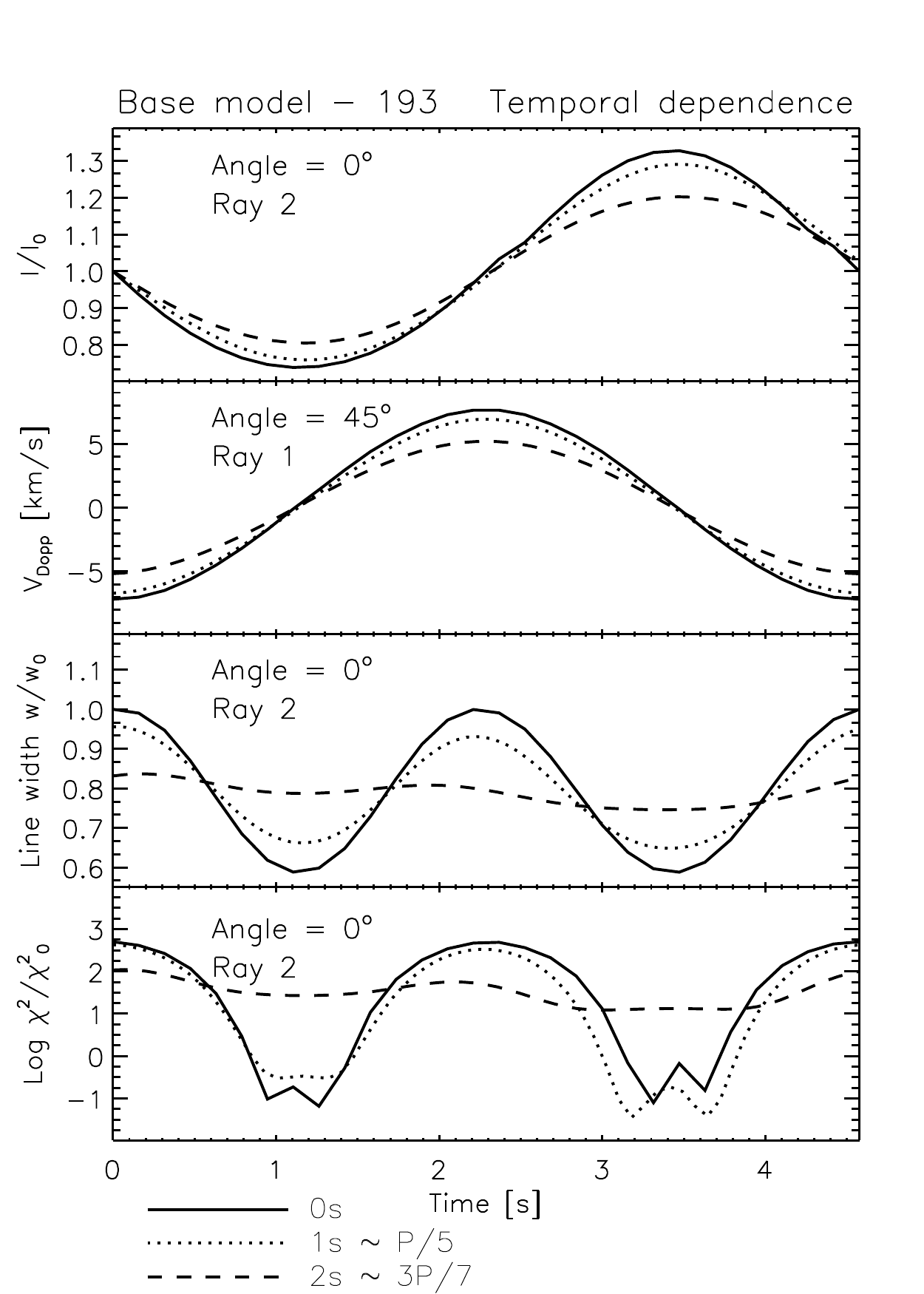}
\end{center}
\caption{Effects of the temporal resolution on the modulation of observable quantities. We take the \textit{base model - 193}, fix the pixel size resolution to 0R and consider cadences of `0s' ($=0.15$~s, solid curve), `1s' ($=1~$s) $\simeq$ P/5 (dotted curve) and `2s' ($=2~$s) $\simeq$ 3P/7 (dashed curve), where P $\simeq4.5$~s is the period of the mode. The upper panel shows the temporal evolution of the normalised integrated intensity along the line-of-sight for ray 2 taking a viewing angle of $0^{\circ}$. The rest of the panels, respectively from top to bottom, show the temporal evolution of the Doppler velocity, the line width and (logarithm of) the $\chi^{2}$ non-Gaussianity parameter resulting from Gaussian fits to the specific intensity profile $I_{\lambda}(t)$ for the different temporal resolutions. In order to show best the effects on the Doppler velocity, only for this quantity we take the case of ray 1 with a viewing angle of $45^{\circ}$. The normalisation factor for the line width is the initial value at time t=0 for the 0s temporal resolution. That of the $\chi^{2}$ parameter is the time average value of the profile for ray 1 for 0s temporal resolution.}
\label{fig13}
\end{figure}
\begin{figure}[!h]
\begin{center}
\includegraphics[width=6cm]{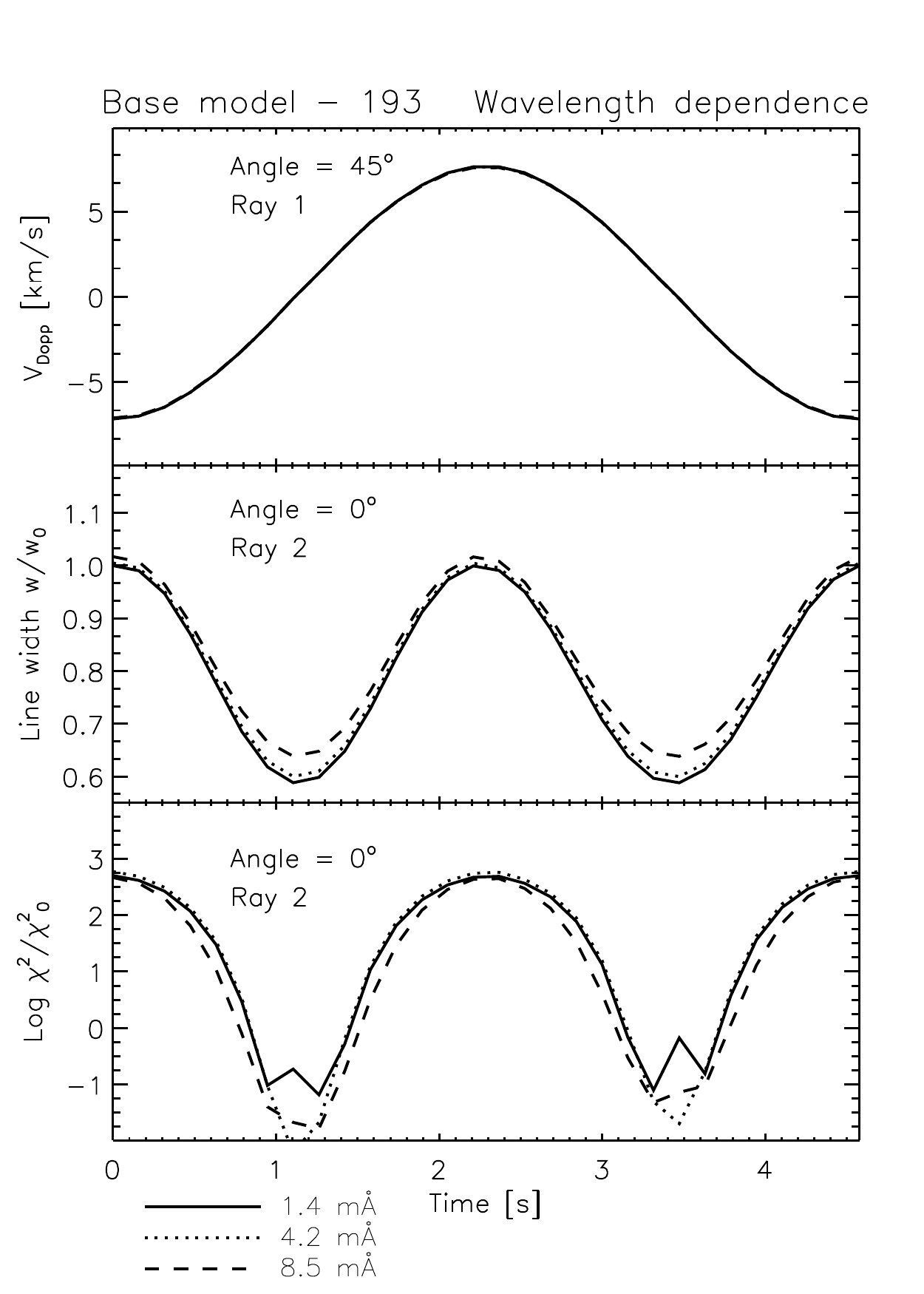}
\end{center}
\caption{Temporal evolution of the Doppler velocity (upper panel), the line width (centre panel) and (logarithm of) the $\chi^{2}$ non-Gaussianity parameter (lower panel) resulting from Gaussian fits to the specific intensity profile $I_{\lambda}(t)$ obtained from the \textit{base model - 193}, for different wavelength resolutions. We consider samplings of 1.4~m$\AA$ (solid curve), 4.2~m$\AA$ (dotted curve) and 8.5~m$\AA$, leading to 48, 16 and 8 points along the spectral line, respectively. In order to show best the effects of the wavelength we fix the pixel size resolution to 0R and show the case of ray 1 with a viewing angle of $45^{\circ}$ for the top panel and the case of ray 2 with a viewing angle of $0^{\circ}$ for the middle and bottom panels. The normalisation values are the same as in Fig.~\ref{fig13}.}
\label{fig14}
\end{figure}

Figure \ref{fig13} shows the effect of temporal resolution on the integrated intensity and spectral broadening obtained from the modulation by the sausage mode for the \textit{base model - 193}. We choose a viewing angle of $0^{\circ}$ and ray 2 (anti-node), for which the effects are strongest for the integrated intensity (top panel), line width (third panel from top to bottom), and $\chi^{2}$ parameter (bottom panel), and a viewing angle of $45^{\circ}$ and ray 1 (node) for the Doppler velocity (second panel from top to bottom). We also fix the spatial resolution to 0R. Different temporal resolutions are obtained by integrating the intensity profiles over temporal windows corresponding to the designated resolution. Since the period of the sausage mode is about 4.5~s only a few points of the full profile are obtained per period. In order to obtain full period profiles approximately 7 and 13 periods are needed, respectively, for 1~s and 2~s resolutions. Longer time integrations also generate phase differences between the profiles equal to the resolutions of each profile. We choose however to correct for this effect and present the shifted profiles in Fig.~\ref{fig13} in order to facilitate comparison.

As seen in the top panel of Fig.~\ref{fig13}, the intensity variation from the sausage mode
decreases with temporal resolution. Factors ranging from 1.2 to 1.5 were found among the models. Although this effect may not appear important for strong intensity variations as those obtained for the \textit{base model - 193}, they are especially relevant for low intensity variations, as those obtained in the \textit{base model - 171}. Longitudinal variation can therefore be erased in this case. 

In the lower 3 panels of Fig.~\ref{fig13} we show the effects of temporal resolution on the Doppler velocity, the line width and the $\chi^{2}$ non-Gaussianity parameter obtained from the Gaussian fits to the specific intensity profiles $I_{\lambda}(t)$. The Doppler velocity, being the displacement of the bulk of the plasma along the line-of-sight, is the least affected quantity, with maxima decreasing from 7~km~s$^{-1}$ to $\lesssim5$~km~s$^{-1}$ (decreasing factors between 1.1 to 1.5 among the models). On the other hand, the strength of the spectral broadening depends mostly on the fast components of the plasma. Correspondingly, the line width variations of about $40\%$ obtained in the model are reduced to $30\%$ and $<10\%$, respectively, for time cadences of 1~s and 2~s, compromising its detectability with the current instruments. Accordingly, decreasing factors between 4.6 and 6.3 are found between the models. Similarly, the variation of the $\chi^{2}$ parameter is reduced (spectra become more and less Gaussian at times of expansion and compression, respectively), especially in the 2~s cadence case, where it is reduced by several orders of magnitude. Hence, for the 2~s time cadence, the spectral line is seen as a profile with almost constant shape and width roughly equal to the average width of the true profile, but for which bulk Doppler motions can still be observed.

\begin{figure*}[!ht]
\begin{center}$
\begin{array}{cc}
\includegraphics[width=8cm]{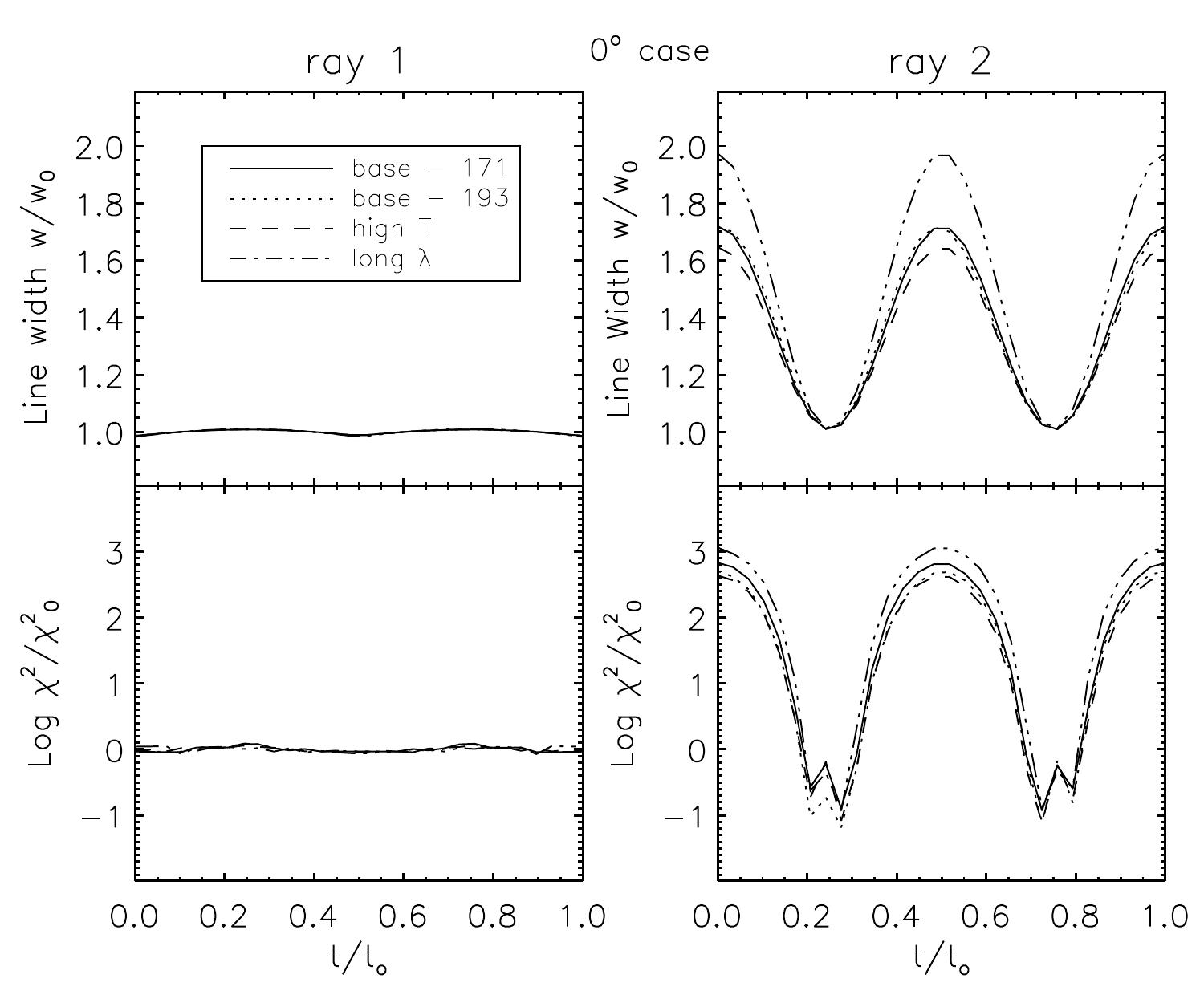} &
\includegraphics[width=8cm]{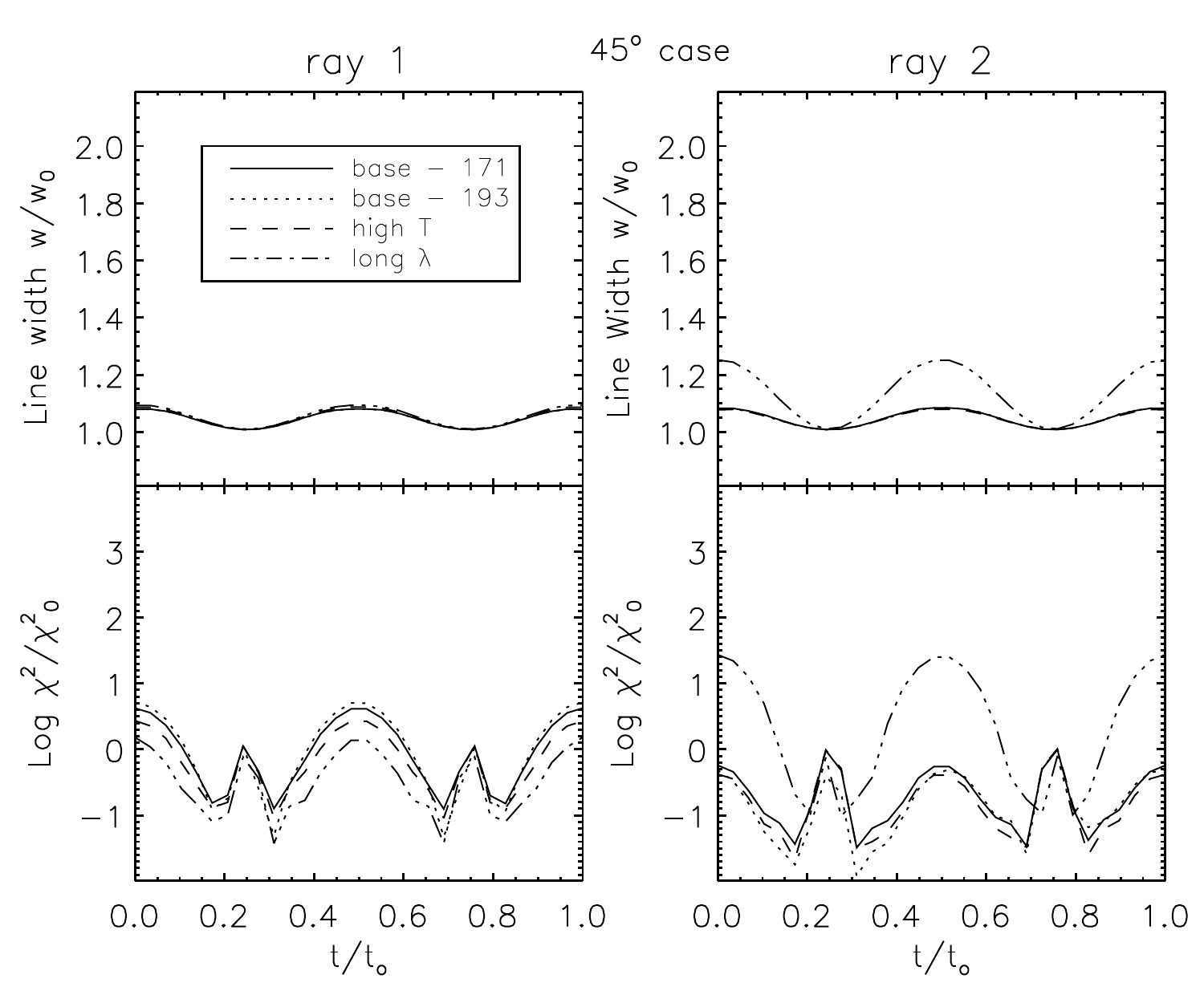}
\end{array}$
\end{center}
\caption{Model comparison of the temporal variation of the spectral line broadening. The line width (upper panels) and (logarithm of) the non-Gaussianity $\chi^{2}$ parameter (lower panels) resulting from the Gaussian fits to the specific intensity $I_{\lambda}(t)$ for the 0R pixel size case are presented. From left to right the first two panels correspond to rays 1 and 2 for the $0^{\circ}$ angle case. The last two panels correspond to rays 1 and 2 for the $45^{\circ}$ angle case. Different line styles denote the different models studied in this work, as indicated in the legend. The normalisation values are the time averages of the corresponding quantities for ray 1 and $0^{\circ}$ viewing angle.}
\label{fig15}
\end{figure*}

The effects of wavelength resolution on the spectral broadening of the specific intensity profile $I_{\lambda}(t)$ for the \textit{base model - 193} are shown in Fig.~\ref{fig14}. As for Fig.~\ref{fig13}, here we show the case of ray 1 (node) with a viewing angle of $45^{\circ}$ for the Doppler velocity and the case of ray 2 (anti-node) with a viewing angle of $0^{\circ}$ for the line width and the $\chi^{2}$ parameter. We fix the spatial resolution to 0R. Different spectral resolutions are produced by integrating the (high wavelength resolution) intensities over specific wavelength bins. We can see that no changes are produced in the Doppler velocity. This is due to the lack of asymmetries in our model: Regardless of the non-Gaussianity of the profiles they will always be symmetric (with respect to the wavelength corresponding to the bulk velocity of the plasma). The effects on the line width are also minimal, with slightly increasing factors between 1.04 and 1.09 among the models. This increase is due to the wider wavelength bins over which intensity is added. At lower resolution the bumps created by the density profile cannot be detected, therefore increasing the Gaussianity of the profile. Factors of increase range between 1.20 and 1.23 between the models.

\subsection{Differences between models}\label{diffmodel}

The results presented in the previous section apply in a qualitative manner to all the models presented in this paper. Apart from the differences in the integrated intensity between the $171~\AA$ line (\textit{base model - 171}) and the $193~\AA$ line (\textit{base model - 193}) treated in section~\ref{imaging}, differences between spectral broadening of the different models were also briefly stated in section~\ref{spectro} and \ref{cad_wav}.

Figure~\ref{fig15} shows the time evolution of the spectral line broadening (the line width and the non-Gaussianity $\chi^{2}$ parameter) resulting from the Gaussian fit to the specific intensity $I_{\lambda}$ for all the models presented in this study, taking 0R as fixed pixel size. From left to right the first two panels correspond to the $0^{\circ}$ angle case, for rays 1 and 2. The last two panels correspond to the $45^{\circ}$ angle case for rays 1 and 2. The quantities are normalised with respect to the temporal average of the corresponding quantity for ray 1 with a $0^{\circ}$ viewing angle. 

As can be seen in the panels containing the line profile width all models except the \textit{long $\lambda$ model} have essentially the same variation. This is mainly due to the fact that the same velocity field is achieved for the same wave number. Longer wavelengths (small wave number) produces higher velocity fields, leading to larger variation in line widths. Higher velocities also lead to stronger compression and rarefaction, and therefore larger variation in temperatures and densities. The effects on the spectral broadening of the line are low, however, since plasmas with a hotter temperature produce lower intensities in the $171~\AA$ line. The longer wavelengths produce wider regions of compression and rarefaction, thus modifying the emission crossed by rays 1 and 2 between the models. This can be observed in the $45^{\circ}$ angle case for ray 2, leading to stronger variation in line widths and $\chi^{2}$ values. This effect disappears for a viewing angle of $60^{\circ}$, where ray 2 crosses mixed regions. On the other hand, the high T model shows in general the lowest values in line width and $\chi^{2}$ parameter values (within models with same wave number especially). This is due to the hot plasma component at rest outside the cylinder. Only minimal differences occur between the \textit{base model - 171} and \textit{base model - 193}. This is because the emission measure weighted radial velocities are very similar between the two models. 

\section{Discussion}\label{discuss}

\subsection{Intensity modulation}

Table~\ref{table1} summarizes the main findings of this work. These results have been obtained assuming equilibrium ionisation for the considered ions, and should be discussed taking this assumption into account. As discussed in section~\ref{imaging}, based on the results in Table~\ref{table2}, in the case of a Maxwellian plasma distribution non-equilibrium ionisation severely affects the modulation of intensity, but does not alter the results obtained for the modulation of spectral line shape (width and Gaussianity). The results for intensity presented in Table~\ref{table1} and discussed in this section constitute therefore upper limits to the more realistic case of plasma modulation in a simulation where the ion balance equations are solved.

The modulation of the integrated intensity, especially important for imaging instruments, severely depends on the selected spectral line(s) of the observation. In the case of a coronal line formed at temperatures below those present in the structure, such as the 171~$\AA$ line for the \textit{base model - 171}, the intensity is only lightly modulated by the sausage mode, even when increasing the mode's wavelength, with maximum amplitudes of $2-4~\%$. Such models present intensity variations in anti-phase with the temperature and density variation, due to the negative slope of the ion population function at these temperatures. However, the modulation can be increased by a factor of $10-15$ when considering coronal lines with formation temperature higher than that of the plasma, as is the case for the \textit{base model - 193}. In this case we have an in-phase variation between these quantities. 

In our model the intensity is governed by the ion population fraction rather than the density of emitting ions. Indeed, as shown by Figure~\ref{fig6}, the change in the contribution function from the temperature modulation (at average density) is one order of magnitude larger than the change from the density modulation (at average temperature). This implies that non-equilibrium ionisation effects will be important, especially for \ion{Fe}{xii}, in which the modulation of intensity would completely disappear for the short timescales of the sausage mode (as discussed in section~\ref{imaging}). The important role of the contribution functions of spectral lines, especially concerning non-equilibrium ionisation effects, have been discussed by \citet{DeMoortel_Bradshaw_2008SoPh..252..101D}. Through synthetic intensity modelling of TRACE channels, the authors show similar results as those stated above. They further show that double periodicities can be obtained from the shape of the response function of the instrument due to perturbations leading to temperatures that drift in and out of the maximum formation temperature of the line.

Variation of the modulation of the intensity by kink and sausage modes due to variation of the angle between the line-of-sight and the emitting structure were first analysed analytically by \citet{Cooper_2003AA...397..765C, Cooper_2003AA...409..325C}. Considering the effect of curvature they showed that propagating kink and sausage modes could produce an increase of intensity along part of the loop, such as that observed by \citet{Williams_etal_2001MNRAS.326..428W,Williams_etal_2002MNRAS.336..747W}. In their work the intensities are approximated by the line-of-sight column thickness $l$ and the density $\rho$, as $I\approx\rho^2l$. As discussed in section~\ref{imaging}, the neglect of the contribution function can be misleading, and the variation of line intensities can be severely underestimated. In the present work we have monotonically decreasing dependence on the line-of-sight angle for the range of parameters considered here, matching the results of \citet{Cooper_2003AA...397..765C}. In \citet{Cooper_2003AA...409..325C} the case of a smooth density profile is considered (with the corresponding dispersion relations associated with that profile), which introduces an optimal angle for which the intensity variations are largest. This is due to the fact that in a 2D model the MHD modes are more compressive, and therefore produce significant density variations outside the cylinder. This effect is however severely reduced for 3D geometries, as shown by \citet{Luna_etal_2008ApJ...676..717L}, suggesting that the existence of an optimal angle could be debated. The effect of such density profiles in a 3D model such as the one considered here is the subject of future research.

\citet{Gruszecki_etal_2012AA...543A..12G} have performed 3D numerical simulations of (trapped and leaky) sausage modes oscillating in a cylinder with a smooth density profile and analysed the modulation on the integrated intensity. As in \citet{Cooper_2003AA...397..765C}, the intensity is approximated by integrating the density squared along the line-of-sight (for a viewing angle of $0^{\circ}$), but they further consider the pixel size resolution. With the same amplitude as used in this work they find intensity variations on the order of 2~\% with 0R pixel size resolution, which are strongly reduced by a factor of roughly 15 when increasing the pixel size to 3R (one wavelength), thus matching the results presented here. As can be seen in the table, among all parameters, we find the pixel size resolution to have a dominant role. A pixel size resolution on the order of the wavelength erases the longitudinal structuring in all models. 

\begin{table*}
\caption{\label{table1}Influence of line-of-sight angle, and spatial, temporal and wavelength resolution on the modulation of observable quantities caused by the sausage mode.}
\centering
\begin{tabular}{c c c c c c c c}
\hline
\hline
\multirow{2}{*}{Quantity} & \multirow{2}{*}{Ray} & \multirow{2}{*}{Amplitude} & Angle & Spatial & Cadence & $\lambda~(\mathrm{m}\AA)$ & \multirow{2}{*}{Period} \\
 & & & $0^{\circ}\to60^{\circ}$ & 0~R$\to3~$R & 0~s$\to2$~s & $1.4~\to8.5$ & \\ \hline
\multirow{2}{*}{Intensity} & 1 & $\sim0$ & $\sim0$ & $\sim0$ & \multirow{2}{*}{$\downarrow1.2-1.5$} & \multirow{2}{*}{--} & \multirow{2}{*}{P} \\  \cline{2-5}
 & 2 & $2-4~\%$ (35~\%, 193) & $\downarrow2.5$ & $\downarrow10-17$  & & & \\
 \hline
 Doppler & 1 & $\lesssim1~$km~s$^{-1}$ & 1 & 1 & \multirow{2}{*}{$\downarrow1.1-1.4$} & \multirow{2}{*}{1} & \multirow{2}{*}{P} \\  \cline{2-5}
 velocity & 2 & 7~km~s$^{-1}$ & $\updownarrow\ast$ & 1 & & & \\
 \hline
  Line & 1 & $7-20~\%$ & $\updownarrow\ast$ & $\uparrow\ast$ & \multirow{2}{*}{$\downarrow4.6-6.3$} & \multirow{2}{*}{$\downarrow1.04-1.09$} & \multirow{2}{*}{$\frac{\mathrm{P}}{2}$} \\  \cline{2-5}
 width & 2 & $38-49~\%$ & $\downarrow5.6-11.3$ & $\downarrow2.5-3.4$ & & & \\
 \hline
  \multirow{2}{*}{$\chi^{2}$} & 1 & $2.1-2.9$ & $\downarrow\ast$ & $\updownarrow\ast$ & \multirow{2}{*}{$\downarrow4-5$} & \multirow{2}{*}{$\uparrow1.20-1.23$} & \multirow{2}{*}{$\frac{\mathrm{P}}{4}, \frac{\mathrm{P}}{2}$} \\  \cline{2-5}
  & 2 & $3.75-4$ & $\downarrow2.1-2.5$ & $\updownarrow2.2-5.5$ & & & \\
 \hline
 \end{tabular}
 \tablefoot{Since the sausage mode introduces longitudinal structuring we show values for rays 1 (node) and 2 (anti-node) for geometrical quantities. The amplitude denotes the maximum variation in time with respect to the value at t=0 of the quantity considering all parameters. The unit corresponds to percentage except for the Doppler velocity where it denotes the maximum value, and the $\chi^{2}$ parameter, where it denotes the variation in orders of magnitude. A range in the values of the quantity corresponds to variations between the models. The values in the other columns denote the factor by which the variation of the quantity is modified when varying the specific parameter of the column. The direction of the parameter variation is indicated in the column head. Upside and downside arrows (or both) denote, respectively, an increase or a decrease (or both if non-monotonic) by the specified factor as the parameter is varied. An asterisk indicates that the variation in time of the quantity is negligible for the initial value of the parameter. Values in parenthesis correspond to some relevant values for the \textit{base model - 193}. The last column corresponds to the possible observed period in the specific observed quantity, with P the period of the sausage mode.}
\end{table*}

The sausage mode in the corona oscillates with relatively short periods. Indeed, in the non-leaky regime and for low-$\beta$ plasma conditions, the period of an MHD sausage mode can be shown to satisfy the relation $P<\frac{2\pi R}{j_0 v_{A_i}}$, where $j_0\approx2.4$ is the first zero of the Bessel function $J_0(x)$ \citep{Nakariakov_2003AA...412L...7N, Kopylova_etal_2007AstL...33..706K}. In our models, the periods oscillate around 5~s, making the cadence of the instrument a very important parameter. Intensity variations can be decreased by a factor of 1.5 and more than 3 to 5 periods of the mode need to be observed in order to half resolve the full profile. 

All the interpretations as sausage modes in observations are mostly based on qualitative arguments, especially restricted by the spatial resolutions of the instrument, which in most cases does not allow to investigate in detail the longitudinal variation of the oscillations along the flaring coronal structure. Still, resolutions such as that of NoRH have allowed the detection of different oscillatory power between loop top and footpoint, and allow more precise identification of the mode harmonics \citep{Nakariakov_2003AA...412L...7N, Melnikov_etal_2005AA...439..727M, Inglis_etal_2008AA...487.1147I, Inglis_Nakariakov_2009AA...493..259I}. In these works, the fundamental mode and sometimes the first overtone can be identified. As discussed in \citet{Andries_etal_2009SSRv..149....3A}, observations of higher overtones are specifically important, since they significantly improve physical parameter estimates from application of MHD seismology. Observations of a fundamental mode at the apex of loops would be comparable to the case of ray 2 in our model, while observations towards the footpoints would correspond better to the case of ray 1. Most observations of sausage modes report short and thick loops, with lengths around 25~Mm. In this case, the resolution of NoRH ($5\arcsec-10\arcsec$) corresponds to a pixel size below $\frac{\ell_0}{7}$ or $\frac{\ell_1}{3.5}$, with $\ell_0$ and $\ell_1$ being the wavelengths of the fundamental mode and the first overtone. According to the results presented here a pixel size of $\frac{\ell_1}{3.5}$ reduces the intensity variations by a factor of $30~\%-40~\%$. For detection of higher harmonics the pixel size starts to severely decrease the intensity variation, as seen in the lower bottom panel of Fig.~\ref{fig5}. Observations of modes with large line-of-sight angles with respect to the normal to the loop plane can easily have rays crossing mixed regions of expansion and contraction, corresponding for instance to nodes of the first overtone and the fundamental mode. These scenarios would bring results that are qualitatively similar as those obtained in the present model when varying the line-of-sight angle.  

\subsection{Spectral line broadening modulation}

Table~\ref{table1} especially shows that even when intensity variations are low (as in the \textit{base model - 171}), spectral line broadening can be important, and could therefore be observed by current spectrometers. In section~\ref{spectro} we have shown that the nature of the spectral line broadening from the sausage mode is not thermal but mainly turbulent broadening. This implies that non-equilibrium ionisation will not influence the obtained spectrometric results.

The sausage mode is essentially a radial mode, and therefore Doppler motions (through single-Gaussian fitting) can only be detected when looking at an angle as shown in section~\ref{spectro}. In this case, bulk motions can be detected as blueshift and redshift excursions below $7~$km~s$^{-1}$, roughly one third of the maximum velocities achieved in the present model. Maximum variations are obtained for $0^{\circ}$ viewing angle for the line width and the $\chi^{2}$ parameter for ray 2 (anti-node), with values close to 50~\% and 4 orders of magnitude, respectively (see table). As shown in Fig.~\ref{fig11} longitudinal structuring can be detected in Doppler velocity, line width and the $\chi^{2}$ parameter for angles up to $30^{\circ}$, pixel size resolutions up to 1R, temporal resolution of 1~s (although with 2~s resolution it is still possible to detect bulk Doppler velocities) and for all considered wavelength resolutions. Furthermore, the line width and $\chi^{2}$ parameter present double periodicities. For lower resolutions or higher viewing angle the $\chi^{2}$ parameter presents quadruple periodicity, but in these cases all quantities exhibit variations decreased by factors of 2 to 10 for each parameter, and therefore may become barely detectable. 

So far very few reports exist of periodic spectral broadenings in EUV lines with little or no associated intensity variations interpreted as fast sausage modes. Most of these studies are performed with the Hinode/EIS imaging spectrometer \citep{Culhane_2007SoPh..243...19C}, with the $1\arcsec$ slit (with a pixel size resolution of $2\arcsec$) and temporal resolutions on the order of 10~s or more. \citet{Kitagawa_Yokoyama_2010ApJ...721..744K} Fourier analyze EUV intensity, Doppler velocity and line width time series of data in an active region observed in the \ion{Fe}{xii}~195 line. Moss regions were found to exhibit significant correlated intensity and Doppler velocity oscillations, with variations of $0.4~\%-5.7~\%$ and $0.2-1.2~$km~s$^{-1}$ respectively, consistent with fast MHD waves modulation. Depending on the significance level of intensity oscillations to Doppler velocity and line width oscillations the waves were interpreted as either fast sausage modes, Alfv\'{e}n or kink modes. However, our results suggest that a low significant level in intensity can be obtained with a sausage mode (any pixel size resolution for 171 line, 3R pixel size resolution for 193), while keeping a relatively high significant level for Doppler velocities and line width (above 10~\% variation). The long reported periods on the order of minutes could be produced by leaky fast sausage modes, especially towards the footpoints of loops, as explained below.

Many reports exist of observations in the same lines having similar attributes, interpreted as slow modes, kink modes or Alfv\'{e}n waves \citep[see review by ][]{DeMoortel_Nakariakov_2012RSPTA.370.3193D}. Although the authors discard the sausage mode interpretation for other reasons, the essential characteristic discarding sausage modes are the long reported periods. Sausage modes with long periods can exist in the corona, as shown by \citet{Pascoe_etal_2007AA...461.1149P}, but are leaky and therefore their observability is severely limited for ranges matching the reported periods. 

Often, observations of propagating disturbances are interpreted as slow modes based on the phase speed closely matching the local sound speed, periods between 2 and 10 minutes, (low) intensity variations (of a few percent with respect to the background intensity) and the intensity fluctuation lagging the Doppler shifts by 1/4 period \citep{Wang_etal_2003AA...406.1105W}. Here we would like to stress the fact that it is the combination of all these characteristics that supports such interpretation. Apart from the first two, (phase speed and period), a sausage mode can also explain the rest (intensity modulation and phase shift, as obtained when observing with a non $0^{\circ}$ angle). 

Wavelength resolution is found to play a minor role, relative to spatial and temporal resolutions. As shown in section~\ref{diffmodel} only minor differences appear in the spectral broadening (and none in the Doppler velocity) when decreasing the wavelength resolution to only 8 points across the spectrum ($\Delta\lambda=$8.5~m$\AA$). This may be due to the symmetric nature of the sausage mode (with respect to the axis of the cylinder), but also to the absence of non-uniformities in density or temperature in the present models.
\citet{Pascoe_Nakariakov_2007SoPh..246..165P, Pascoe_Nakariakov_2009AA...494.1119P} considered the effects of varying loop cross-section and fine multilayered structuring on the resonant periods and the spatial profiles of fast sausage mode oscillations. These numerical studies find that the resonant properties of long-wavelength sausage standing modes of the slab are not sensitive to the details of fine structuring and the spatial profiles are only weakly affected by the cross-section inhomogeneity. This suggests that the inherent symmetries of the fast sausage mode are rather robust to loop divergence and internal density fine-structuring, and therefore that the properties of the modulation found for simple cylindrical geometries, such as those used here, are robust as well. 

As discussed in the previous section, the detection of sausage modes is severely limited by its leaky nature in most coronal conditions. On the other hand the lack of reported observations of sausage modes with spectrometers may also be due to an instrumental reason. Our results show that high spatial and temporal resolution in EUV lines is essential for the detection of such waves. Perhaps the most promising instrument for the detection of these waves in the near future is IRIS, the Interface Region Imaging Spectrograph, whose launch is scheduled for the spring 2013. IRIS is specially designed for spectrometric observations of the solar atmosphere in EUV lines with high spatial and temporal resolutions. For instance, observation in the \ion{Fe}{xii}~1349~$\AA$ line ($\log T=6.2~\mathrm{K}$) will have a cadence of $1-2~$s, a spatial resolution of $0.4\arcsec$ (with a field of view of $0.3\arcsec\times40\arcsec$), an effective area of 2.8~cm$^2$ and a wavelength resolution of $\Delta\lambda=12.5~$m$\AA$. 

The new Atacama Large Millimeter/submillimeter Array (ALMA) telescope has started to operate and offers unique advantages for solar observations. Spatial resolutions from $0.06\arcsec$ at 85~GHz to $0.005\arcsec$ at 950~GHz and short integration times of 32~ms are expected, although for solar observations a special filter is needed, which may considerably reduce the capabilities of the telescope. Nevertheless, flare and chromospheric research may considerably benefit from ALMA, especially in the subject of fast waves such as the ones studied here.

The results obtained here would correspond to the ideal case of an imaging spectrometer with at least 2~s cadence, a pixel size resolution of at least one wavelength, and a wavelength resolution of at least 8 points across the line. Furthermore no blending of lines emission was considered, as is produced when observing with the channels of the AIA instrument on board the Solar Dynamics Observatory, no background optically thick contribution and no instrumental noise. Together with the effects of curvature and non uniform media (leading to different wave dispersion relations) these scenarios will be treated in future work. 

\section{Conclusions}\label{conclus}

In this paper we aimed to understand the variations of observable quantities introduced by the fast MHD sausage mode in coronal structures, considering geometrical and instrumental effects. Complexities introduced by non-uniformity of the cylinder or the background, curved geometries, or complex instrumental effects such as instrumental noise are not taken into account. Although important for realistic models of loops, such characteristics are secondary and their effects on intensity modulation cannot be properly tackled before understanding the effects of more simple parameters on simple geometrical models such as the ones considered in this work.

Details from the main results of this work can be found in Table~\ref{table1}. The main findings can be summarized as follows. 

The intensity modulation is strongly dependent on the formation temperature of the spectral line with respect to the temperature of the plasma. Assuming equilibrium ionisation, for plasmas with higher temperature than the formation temperature of the line, such as the case of the \textit{base model - 171}, the modulation is low ($2~\%-4~\%$) and is slightly affected by the geometrical variation of the cylinder. On the other hand, for plasmas with lower temperature than the formation temperature of the line, such as the \textit{base model - 193}, the modulation matches the thermodynamical evolution (density and temperature), and can therefore be significant ($35~\%$). Non-equilibrium ionisation would completely eliminate the \ion{Fe}{xii} intensity modulation (and reduce by a factor of 1.5 or more the \ion{Fe}{ix} intensity) due to the fast timescales of the sausage mode. Simple approximations to the intensity taking into account the column thickness and disregarding the contribution function $G_{\lambda_{0}}(T,n_e)$ can therefore be very misleading.

Depending on the contribution function, the intensity can be in anti-phase with the density (or temperature) profile (case of the \textit{base model - 171}) or in-phase (case of the \textit{base model - 193}). The line width and the $\chi^2$ parameter are in phase with the square of the radial velocity. The $\chi^2$ parameter can also respond to the density variation, which becomes significant for coarse pixel size or large viewing angle. The Doppler velocity is in anti-phase with the radial velocity when watching at an angle (non $0^{\circ}$). Correspondingly, for these angles, the Doppler velocity presents a phase shift of one fourth of the period with the intensity. In accordance with the previous result, the intensity and Doppler velocity exhibit the same periodicity as the mode. Double periodicities are characteristic of the line width and the $\chi^2$ parameter. The $\chi^2$ parameter can present quadruple periodicities for coarse pixel size or large viewing angle. 

The detection of intensity modulation from sausage modes is strongly dependent on the angle of observation. The intensity decreases with the viewing angle (where $0^{\circ}$ angle denotes the perpendicular to the axis of the cylinder) by a factor of 2.5 for the studied range. Furthermore, the intensity modulation can vary significantly longitudinally (along the axis of the cylinder). Rays crossing the cylinder at nodes exhibit considerably less variation than those crossing the anti-nodes (a factor of 20 or more, depending on the pixel size resolution).

Setting aside equilibrium ionisation effects, the detection of intensity modulation is strongly dependent on spatial and temporal resolutions. Both can be the dominant parameters, especially if the pixel size is on the order of the mode's wavelength or the temporal resolution is on the order of the mode's period. A pixel size or a cadence on these orders (respectively) can completely erase the longitudinal structuring, therefore eliminating the possibility of detection. Wavelength resolution is not particularly important for the case of sausage modulation. This is due to the inherent symmetry of the sausage mode with respect to the axis of the cylinder, leading to symmetric line profiles. Numerical studies suggest that this as well as most results presented here may be robust to the presence of inhomogeneities in the model (leading to possible assymmetries), as well as effects from varying cross-section along the cylinder. 

Even when the intensity modulation is low, the variation of the Doppler velocity, and especially the line width and the $\chi^2$ non-Gaussianity parameter can be significant (a factor of up to $40~\%$ for line width and up to four orders of magnitude change for the $\chi^2$ parameter), placing spectrometers in a clear advantage over imaging instruments for the detection of these waves. This result is independent of the ionisation state of the plasma since the spectral line is mainly affected by the unresolved turbulent motions rather than the thermal motions. Significant variability in all quantities can still be obtained when viewing at an angle of up to $30^{\circ}$ (Doppler velocities can especially be detected for the range $15^{\circ}-45^{\circ}$), with pixel size resolutions up to 1R, or temporal resolution of 1~s (although with 2~s resolution it is still possible to detect significant Doppler velocities) and for all considered wavelength resolutions. We consider this to be the most important result of this paper.

\begin{acknowledgements}
The authors would like to thank the referee for very constructive comments and Dr. Verwichte and Dr. De Moortel for productive discussions, all leading to a significant improvement of the manuscript. PA further thanks Dr. D. Jess for allowing use of his Irish monster computer for the high resolution calculations. TVD and PA have received funding from the Odysseus programme of the FWO-Vlaanderen. TVD also acknowledges funding from the EU’s Framework Programme 7 as an ERG with grant number 276808. CHIANTI is a collaborative project involving the NRL (USA), the Universities of Florence (Italy) and Cambridge (UK), and George Mason University (USA).
\end{acknowledgements}

\bibliographystyle{aa}
\bibliography{ms.bbl}  

\clearpage

\end{document}